\documentclass[10pt,journal,compsoc]{IEEEtran}
\usepackage{booktabs}
\usepackage{colortbl}
\usepackage{color}
\usepackage[table]{xcolor}
\usepackage{xspace}
\usepackage{url}
\usepackage{balance}
\usepackage{tabulary}
\usepackage{array}
\usepackage{graphicx}
\usepackage{hyperref}
\usepackage[ruled,lined,linesnumbered,vlined,algo2e]{algorithm2e}
\usepackage{listings}
\usepackage{lipsum}
\usepackage{hyperref}
\usepackage{graphicx}
\usepackage{amsmath,amssymb}
\usepackage{subcaption}
\usepackage[T1]{fontenc}
\usepackage{amsmath, amsthm, amssymb}
\usepackage[ansinew]{inputenc}
\usepackage[many]{tcolorbox}
\usepackage{xcolor}
\usepackage{extarrows}
\usepackage{multirow}
\usepackage{adjustbox}
\usepackage{colortbl}
\usepackage{tabularx}
\usepackage{enumitem}
\setitemize[0]{leftmargin=10pt}

\newcommand{\oldtool}{{{StoryDroid}}\xspace}
\newcommand{\tool}{{{StoryDistiller}}\xspace}

\newcommand{\revise}[1]{{\textcolor{black}{#1}}}
\newcommand{\revised}[1]{{\textcolor{black}{#1}}}
\newcommand{\rTwo}[1]{{\textcolor{black}{#1}}}
\newcommand{\rThree}[1]{{\textcolor{black}{#1}}}

\usepackage{listings}
\usepackage{courier}
\definecolor{grey}{rgb}{0.52,0.52,0.51}
\definecolor{arsenicgrey}{rgb}{0.23, 0.27, 0.29}
\definecolor{cerulean}{rgb}{0.0, 0.48, 0.65}
\definecolor{cardinal}{rgb}{0.77, 0.12, 0.23}
\definecolor{dartmouthgreen}{rgb}{0.05, 0.5, 0.06}

\SetCommentSty{mycommfont}

\lstdefinestyle{mystyle}{
    commentstyle = \color{dartmouthgreen}, 
    keywordstyle = \color{cardinal},
    stringstyle = \color{magenta}, 
    rulecolor = \color{black},
 	emph={replace,setAdapter,PrefEditor,InnerClass,startActivity,SearchPanel,PartList},
 	emphstyle={\color{cerulean}},
    numberstyle=\tiny\color{arsenicgrey},
    basicstyle=\ttfamily\scriptsize\bf,
    breakatwhitespace=false,
    breaklines=true,
    captionpos=b,
    keepspaces=true,
    numbers=left,
    numbersep=5pt,
    showspaces=false,
    showstringspaces=false,
    showtabs=false,
    tabsize=2,
    frame=tb,
    frameround=fttt,
    xleftmargin=10pt,
    xrightmargin=10pt,
    keepspaces=true,
    language=java
}
\usepackage{url}

\makeatletter
\newcommand*\bigcdot{\mathpalette\bigcdot@{1}}
\newcommand*\bigcdot@[2]{\mathbin{\vcenter{\hbox{\scalebox{#2}{$\m@th#1\bullet$}}}}}
\makeatother

\begin{document}
\title{Automatically Distilling Storyboard with Rich Features for Android Apps}

\author{Sen Chen,
        Lingling Fan, 
        Chunyang Chen,
        and~Yang Liu
\IEEEcompsocitemizethanks{
	\IEEEcompsocthanksitem
	Sen Chen is with the College of Intelligence and Computing, Tianjin University, China. Email: senchen@tju.edu.cn. 
	Lingling Fan (Corresponding author) is with the College of Cyber Science, Nankai University, China. Email: linglingfan@nankai.edu.cn.
	Chunyang Chen is with the Faculty of Information Technology, Monash University, Australia. Email: chunyang.chen@monash.edu.
	Yang Liu is with the Zhejiang Sci-Tech University, China and School of Computer Science and Engineering, Nanyang Technological University.
	Email: yangliu@ntu.edu.sg.
	\IEEEcompsocthanksitem
	\revised{This work is an extension to our previous paper published in ICSE'19~\cite{chen2019storydroid}. This journal version has substantially extended our conference version in terms of technique contributions and experiments.
	}
}}

\markboth{Journal of \LaTeX\ Class Files,~Vol.~xx, No.~xx, xx~2020}%
{Chen \MakeLowercase{\textit{et al.}}: \tool: Automatically Distilling Storyboard with Rich Features of Android Apps}

\IEEEtitleabstractindextext{
\begin{abstract}
Before developing a new mobile app, the development team usually endeavors painstaking efforts to review many existing apps with similar purposes. The review process is crucial in the sense that it reduces market risks and provides inspirations for app development. However, manual exploration of hundreds of existing apps by different roles (e.g., product manager, UI/UX designer, developer, and tester) can be ineffective. For example, it is difficult to completely explore all the functionalities of the app from different aspects including design, implementation, and testing in a short period of time. However, existing reverse engineering tools only provide basic features such as AndroidManifest.xml and Java source files for users.

Following the conception of storyboard in movie production, we propose a system, named \tool, to automatically generate the storyboards for Android apps with rich features through reverse engineering, and assist different roles to review and analyze apps effectively and efficiently. \revise{Specifically, we (1) propose a hybrid method to extract a relatively complete Activity transition graph (ATG), that is, it first extracts the ATG of Android apps through static analysis method first, and further leverages dynamic component exploration to augment ATG; (2) extract the required inter-component communication (ICC) data of each target Activity by leveraging static data-flow analysis and renders UI pages dynamically by using app instrumentation together with the extracted required ICC data; (3) obtain rich features including comprehensive ATG with rendered UI pages, semantic activity names, corresponding logic and layout code, etc. (4) implement the storyboard visualization as a web service with the rendered UI pages and the corresponding rich features. Our experiments unveil that \tool is effective and indeed useful to assist app exploration and review. We also conduct a comprehensive comparison study to demonstrate better performance over IC3, Gator, Stoat, and StoryDroid.}
\end{abstract}

\begin{IEEEkeywords}
Android apps, app review, competitive analysis, reverse engineering, storyboard
\end{IEEEkeywords}
}

\maketitle

\IEEEdisplaynontitleabstractindextext
\IEEEpeerreviewmaketitle

\IEEEraisesectionheading{\section{Introduction}\label{sec:introduction}}
\IEEEPARstart{M}obile applications (apps) now have become the most popular way of accessing the Internet as well as performing daily tasks, e.g., reading, shopping, banking, and chatting~\cite{web:mobileDesktop}. Different from traditional desktop applications, mobile apps are typically developed under the time-to-market pressure and facing fierce competitions --- over 3.8 million Android apps and 2 million iPhone apps are striving to gain users on Google Play and Apple App Store, the two primary mobile app markets~\cite{web:appNumber}. Additionally, a large number of mobile apps still suffer from functional bugs~\cite{fan18, fan18efficiently}, security vulnerabilities~\cite{chen2018ausera,chen2019ausera, chen2018mobile}, and the lack of marketing competitiveness.

Therefore, for app developers and companies, it is crucial to perform extensive competitive analysis through app review over existing apps with similar purposes~\cite{guo2017automated,arbon2014app,web:competitorAnalysis,fox2017mobile}. This analysis helps understand the competitors' strengths and weaknesses, and reduces market risks before development. Specifically, it identifies common app features, design choices, and potential customers. Moreover, researching similar apps also helps developers gain more insights on the actual implementation, given that delivering commercial apps can be time-consuming and expensive~\cite{web:appCost}. Besides, from the {perspective} of app testers for testing purpose, they aim to catch more useful features, such as logic, functionalities, and version changes. However, to the best of our knowledge, existing reverse engineering tools can only provide partial features, such as the configuration file (i.e., AndroidManifest.xml) and Java {source} files to analysts directly~\cite{arnatovich2018comparison}.

To achieve the aforementioned tasks such as competitive analysis, a freelance developer or a product manager (PM) in a tech company has to download the apps from markets, install them on mobile devices, and use them back-and-forth to identify what he is interested in{~\cite{guo2017automated,arbon2014app,fox2017mobile,web:competitorAnalysis}}. However, such manual exploration can be painstaking and ineffective. For example, if a tech company plans to develop a social media app, over 200 similar apps on Google Play will be under review. It is overwhelming to manually analyze them --- register accounts, feed specific inputs if required, and record necessary information (e.g., what are the main features, how are the app pages connected). Additionally, commercial apps can be too complex to be manually uncovered all functionalities in a reasonable time~\cite{azim2013targeted}. For UI/UX designers, the same exploration problem still remains when they want to get inspiration from similar apps' design. In addition, the large number of user interface (UI) screens within the app also makes it difficult for designers to understand the relation and navigation between pages. For developers who want to get inspiration from similar apps, it is difficult to link the UI screens with the corresponding implementation code --- the code can be separated in layout files as well as a large piece of functional code. For app testers who want to understand the existing apps in depth from multiple aspects, such as app logic, functionalities, and version changes in order to design test cases or testing strategies, it is difficult to obtain all the useful features at the same time with existing reverse engineering tools, such as {ApkTool}~\cite{apktool} and {Androguard}~\cite{androguard}.

\begin{figure}
	\center
	\includegraphics[width=0.425\textwidth]{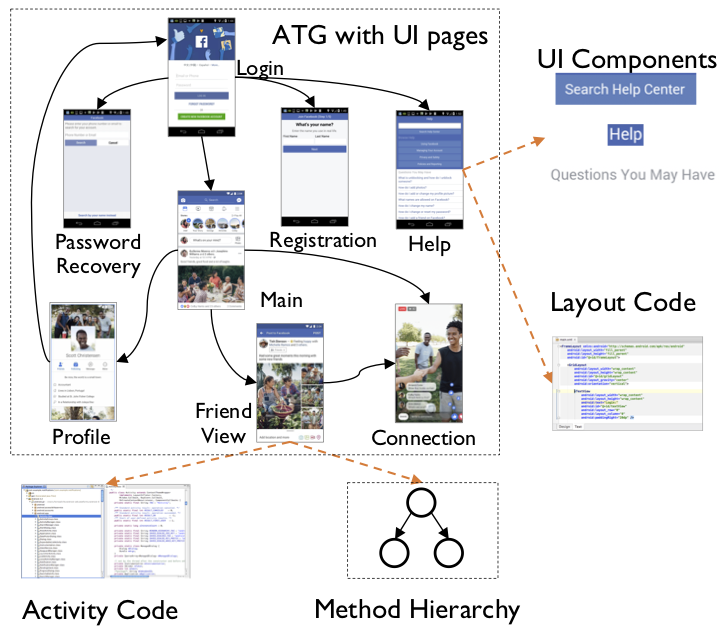}
	\caption{The storyboard diagram of an Android app}
	\label{fig:demo}
\end{figure}

Inspired by the conception of \emph{\textbf{storyboard}}\footnote{``Storyboard'' was developed at Walt Disney Productions, including a sequence of drawings typically with some directions and dialogues, representing the shots planned for a movie or television production.} in movie industry~\cite{finch1995art}, we intend to generate the storyboard of an Android app to visualize its key app behaviors and rich features. Specifically, we use activities (i.e., UI screens) to characterize the ``scenes'' in the storyboard, since activities represent the intuitive impression of the apps in a full-screen window and are the most frequently used components for user interactions~\cite{activity}. Fig.~\ref{fig:demo} shows the storyboard diagram of \emph{Facebook} (one of the most popular social media apps), which includes the activity transition graph (ATG) with UI pages, the detailed layout code, independent UI components, the functional code of each activity (\emph{Activity Code}), and method call relations within each activity (\emph{Method Hierarchy}). Based on this storyboard, PMs can review a number of apps in a short period of time and propose more competitive features in their own app.\footnote{The main purpose is to help PMs, developers, designers, and testers understand and get inspiration from existing apps, instead of directly distributing any part of the code for developing apps for commercial purpose.} UI designers can obtain the most related UI pages for reference. And developers can directly refer to the related code to improve development efficiency. Meanwhile, app testers can understand the main logic, functionalities, as well as version changes to generate test cases.

However, generating storyboards is challenging. \revise{First, ATG is usually incomplete with low activity coverage due to the limitations of static analysis tools such as A3E~\cite{azim2013targeted}, IC3~\cite{octeau2015composite}, and Gator~\cite{gator}.} Second, to render all UI pages, a pure static approach may miss parts of UIs that are dynamically rendered and reduce UI similarity compared with real pages, whereas existing pure dynamic approaches~\cite{monkey, su2017guided, chen2018ui, chen2019codegeneration} can only reach limited activities in the app, especially for those requiring login. Third, the obfuscated activity names lack the semantics of corresponding functionalities, making the storyboard hard to understand.

In our previous conference version~\cite{chen2019storydroid}, to overcome these challenges, we propose a system (named \textbf{StoryDroid}) to automatically generate the storyboards of apps in three main phases: (1) \emph{Activity transition extraction}, which extracts ATG from the apks, especially the transitions in fragments~\cite{fragment} (components of Activity) and inner classes~\cite{inner}, making ATG more complete. (2) \emph{Static UI page rendering}, which first extracts the dynamic components (if any) for each UI page and embeds them into the corresponding static layout. It then renders each UI page statically based on the static layout files. (3) \emph{Semantic name inferring}, which infers the semantic names for the obfuscated activity names by comparing the layout hierarchy with the ones in our database.\rThree{\footnote{\rThree{According to a pilot study on 1,000 randomly selected activities names, we found that \textit{few activity names} lack semantics in the experimental dataset. Therefore, in this version, to make the paper more compact, we did not pay more attention to the \textit{semantic name inferring}.}}} 

\revise{However, there are still some limitations in \oldtool~\cite{chen2019storydroid}, which motivates us to extend to this journal version. (1) The completeness of ATG 
is still not satisfying (below 70\% activity coverage on average) especially for the closed-source apps (below 60\%) due to the limitations of pure static analysis such as decompilation errors and dynamic-loading components.
(2) Some of the rendered pages by the pure static method have a big visual difference compared with the real pages (Fig.~\ref{fig:two_versions} (a)) \rTwo{even though} it achieves \textasciitilde80\% similarity on average. More importantly, not all the dynamic/hybrid layout code can be transferred to static layout code, causing unexpected errors such as rendering failures of user-defined components, third-party dependency errors, 
and resource file errors, which directly leads to low success rate of page rendering (\textasciitilde 55\% launch ratio on average).} 
\revise{These above issues significantly reduce the usability of storyboards in practice.}

\revise{However, it is non-trivial to overcome the above limitations, because it is challenging to further improve (1) the completeness of ATG only by a pure static method because it is hard to handle various types of activity startups or address the limitations caused by code reverse engineering~\cite{octeau2015composite, gator, chen2019storydroid,azim2013targeted}; (2) the capability of static UI page rendering because it cannot transfer various types of dynamic components and is hard to render UI pages of closed-source apps due to compilation failures.
To address the limitations of the pure static method in \oldtool~\cite{chen2019storydroid}, we propose a \textit{hybrid approach} named \textbf{\tool}, which combines static and dynamic methods to distill and generate storyboards for Android apps more effectively, and further help different stakeholders to explore and review apps. Consequently, in this paper, we make substantial effort to upgrade the generation capability of storyboards for apps from the following technical aspects:}

\begin{itemize}
	\item \revise{In terms of the \textit{Activity transition extraction}, we leverage \textit{Dynamic UI component exploration} to dynamically augment the transition graph extracted by the pure static method in \oldtool. Consequently, StoryDistiller combines the advantages of static and dynamic methods with over 20\% increase in activity transition pairs and more than 10\% improvement in activity coverage.}
	
	\item \revise{As for the \textit{Dynamic UI page rendering}, we leverage {static data-flow analysis} to extract the inter-component communication (ICC) data transferred across different activities. Based on it, \tool can render UI pages dynamically with a high success launch ratio (\textbf{\textasciitilde80\%} vs. \textasciitilde 55\% in \oldtool on average) and can address the low page similarity of the static rendering method\rThree{\footnote{\rThree{For the pure static method in \oldtool, rendering the page is based on the static layout files or the transferred layout files for the dynamic/hybrid layouts, and no more other parameters are needed like ICC data using in \tool.}}} used in \oldtool (\textbf{\textasciitilde95\%} vs. \textasciitilde80\% on average).}

	\item \revise{\tool provides a web service to visualize the storyboards with rich features and enhance the usability of \tool. Thanks to the capability of \tool and large-scale dataset of apps, we are able to build a large and multi-dimension dataset with different kinds of data to enable different follow-up research directions.}
\end{itemize}

\revise{Specifically, in this extension, we evaluate \tool on 150 apps (75 open-source and 75 closed-source apps) from the following two aspects}: effectiveness evaluation of each phase of the proposed method and usefulness evaluation of the visualization outputs as a web service. \revise{The experimental results show that (1) for activity transitions, \tool outperforms the existing static methods such as IC3, Gator, and \oldtool (7.8, 10.0, and 18.2 vs. \textbf{23.3} transition pairs on average); For activity coverage, \tool also performs the best compared with the above three static methods (38.7\%, 33.7\%, and 69.6\% vs. \textbf{77.5\%} on average) and the dynamic method (i.e., Stoat) (36.3\% vs. \textbf{77.5\%}). (2) \tool achieves around \textbf{80\%} launch ratio of activities for each app on average on the 150 selected apps, while \oldtool only launches about 55\% activities due to the limitations of the pure static rendering method. 
Moreover, our rendered UI pages clearly show the actual functionalities of the activities compared with the ones that are obtained by manual exploration and achieve over 95\% UI similarity. In addition, the user study shows that with the help of \tool, activity coverage has a significant improvement compared with exploration without \tool when exploring and reviewing apps.}

In summary, we make the following main contributions:

\begin{itemize}
	\item This research work aims to automatically generate the storyboards of Android apps. It assists app development teams including PMs, designers, developers, and app testers to quickly have a clear overview of other similar apps and target different tasks such as app exploration and app review. 
	
	\item \revise{We leverage a hybrid approach to extract a comprehensive ATG for Android apps, and render UI pages dynamically 
	with high UI similarity compared with the real ones.} 
	
	\item \revise{We propose a novel method to render UI pages by obtaining the required ICC data for launching each activity, minimizing unexpected errors when rendering UI pages (Algorithm~\ref{algo:iccdata} in \S~\ref{subsec:dynamic}).}
	
	\item Our comprehensive experiments demonstrate not only the effectiveness of the generated storyboards, but also the usefulness of our StoryDistiller with the extracted rich features for assisting app review and analysis.
	
	\item To enhance the usability of \tool, we visualize the storyboards with all rich features through a web service (Fig.~\ref{fig:service}). We also construct {a multi-dimension} dataset with different kinds of features based on \tool and enable several follow-up research directions, such as extracting commonalities across apps, recommending UI design and code, and guiding app testing. We will gradually release these datasets to enable different research applications~\cite{storydistiller}.
	
	\item \rTwo{We released the code of \tool on GitHub for the community to facilitate the following works: \url{https://github.com/tjusenchen/storydistiller}}

\end{itemize}


\section{Motivating Scenario}\label{sec:motivation}
We detail the typical app review process{~\cite{devprocess,typical,guo2017automated,arbon2014app,fox2017mobile}} with our \tool for Android apps in terms of different roles in the development team. Eve is a PM of an IT company. Her team plans to develop an Android social app. In order to improve the competitiveness of the designed app, she searches hundreds of similar apps (e.g., Facebook, Instagram, and Twitter) based on the input keywords (e.g., social and chat) from Google Play Store. She then inputs all of the URLs of these apps into \tool which automatically download all of these apps with Google Play API~\cite{api}. \tool further generates the storyboard (e.g., Fig~\ref{fig:demo}) of all these apps and displays them to Eve for an overview. By observing these storyboards together, she easily understands the storyline of these apps, and spots the common features among these apps such as registering, searching, setting, user profile, posting, etc. Based on these common features, Eve comes up with some unique features which can distinguish their own app from existing ones.

Alice, as a UI/UX designer, needs to design the UI pages according to Eve's requirements. With our \tool, she can easily get not only a clear overview of the UI design style of related apps, but also interaction relations among different screens within the app. Then, Alice can develop the UI and user interaction of her app inspired by others' apps~\cite{web:designer1, web:designer2}.

Bob is an Android developer who needs to develop the corresponding app based on Alice's UI design. Based on Alice's referred UI design in the existing app, he can also refer to that app with the help of our \tool. By clicking the UI screen of each activity in the storyboard, \tool returns the corresponding UI implementation code no matter it is implemented with pure static code, dynamic code, or hybrid ones. To implement their own UI design, he can refer to the implemented code and customize it based on their requirement. That development process is much faster than starting from scratch. In addition, Bob may also be interested in certain functionality within a certain app. By using \tool, he can easily locate the logic code.

Mallory is an Android tester who has to test the corresponding app based on Bob's implementation. By exploring \tool, she can understand the main logic and functionalities to generate test cases. For apps with multiple versions, \tool is able to identify the UI components that have been modified between different app versions. Therefore, she can also reuse most of the test cases since different versions of a single app have many common functionalities. Reusing test cases is useful to improve the efficiency of app testing.

\section{Preliminaries}\label{sec:background}
In this section, we briefly introduce the concept of Android Activity and Fragment, and the mechanism of inter-component communication (ICC).

\subsection{Android Activity and Fragment}
There are 4 types of components in Android apps (i.e., Activity, Service, Broadcast, and Receiver). Activity~\cite{activity} and Fragment~\cite{fragment} render the user interface and are the visible parts of apps. Activity is a fundamental component for drawing the screens which users can interact with. Fragment represents a portion of UIs in the activities, which contributes their own UI to certain activities. Fragment always depends on an Activity and cannot exist independently. A Fragment can also be reused in multiple activities and an activity may contain multiple fragments based on the screen size, with which we can create multi-panel UIs to adapt to mobile devices with different screen sizes. 
\revise{Service is another important component of Android that is used to perform operations on the background such as playing music and handling network transactions. It does not has any UI.}

\subsection{Inter-component Communication}\label{sec:background-icc}
When an app intends to make inter-component communication (ICC), e.g., start a new activity $B$ or connect to other apps from the current activity $A$, it requires to create an ``Intent'' object describing the task.	If there is other data/messages required to be transferred from activity $A$ to activity $B$, the parameters, such as action, category, and extra parameters can be stored in the ``Intent'' object or in the ``Bundle'' class, and transferred to activity $B$ for successful launching. When activity $B$ receives it, it can parse the data inside and use them to render the UI screen or conduct other transactions. If activity $B$ does not receive the necessary data or the proper form of the necessary data, it could not be rendered successfully, sometimes even causing an app crash, usually ``NullPointerException''. Note that, one activity also can be started by other activities via Fragments and inner classes~\cite{inner}.

As shown in Table~\ref{tbl:iccdata}, the ICC data transferred between components are classified into two categories: \textit{primitive attribute} and \textit{extra parameter}. The primitive attributes are usually stated in the intent-filter element of an activity in the AndroidManifest.xml file, indicating that only the intents with specific attributes can launch the activity, such as actions to be performed (e.g., {android.intent.action.VIEW}), URI data to be operated on (e.g., {vnd.android.cursor.dir/vnd.google.note}), and special flags associated with the ``Intent'', etc. Primitive attributes can also be declared in the Java files. The extra parameters are usually declared in the Java files in an Intent object or Bundle class, indicating the data transferring to the target activity, which is also the necessary data to launch the target activity, in the form of <$key$, $type$> pairs where $key$ is a String \rTwo{indicating the parameter name} and $type$ indicates the data type of the value. \rTwo{For example, if an activity requires a specific ``pid'' (e.g., pid = 2) to be successfully launched, then the $key$ refers to the parameter name ``pid'', and the $type$ refers to data type of 2 (i.e., Integer).}

\begin{figure*}
	\centering
	\includegraphics[width=1\textwidth]{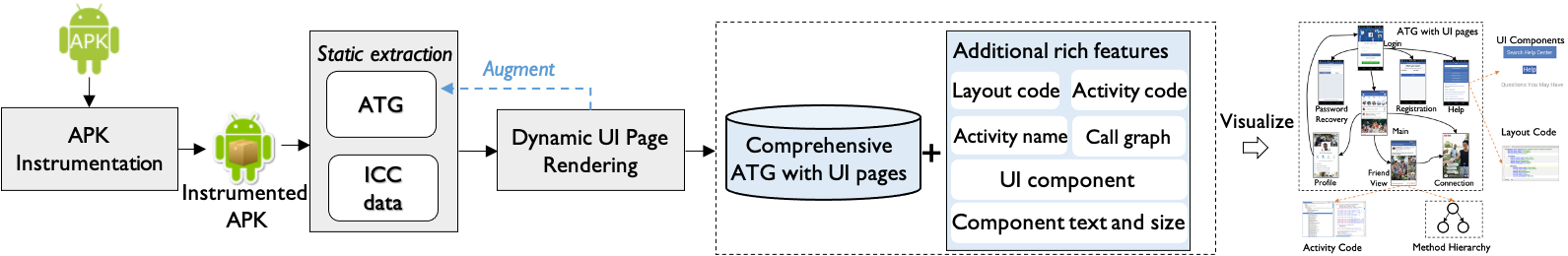}
	\caption{\revise{\rTwo{Overview of \tool}}}
	\label{fig:workflow}
\end{figure*}

\begin{table}[t]
	\centering
	\caption{\revised{Data types transferred in ICC}}
	\label{tbl:iccdata}
	\scalebox{0.95}{\revised{\begin{tabular}{c|c|l}
		\hline
		\textbf{Category} & \textbf{SubCategory} & \textbf{Data Type/Description} \\ \hline
		\multirow{4}{*}{\textit{\begin{tabular}[c]{@{}c@{}}Primitive \\ Attributes\end{tabular}}} & Action & String \\ \cline{2-3} 
		& Category & Set\textless{}String\textgreater{} \\ \cline{2-3} 
		& Data & String \\ \cline{2-3} 
		& Type & String \\ \hline
		\multirow{3}{*}{\textit{\begin{tabular}[c]{@{}c@{}}Extra \\ Parameters\end{tabular}}} & Basic & \begin{tabular}[c]{@{}l@{}}\textless{}$key, type$\textgreater pair, \rTwo{where $key$ refers to the}\\ \rTwo{parameter name, and $type$ indicates the} \\\rTwo{data type of the value (e.g., Integer, String).}
		\end{tabular} \\ \cline{2-3} 
		& Bundle & \begin{tabular}[c]{@{}l@{}}Set of \textless{}$key, type$\textgreater pairs, each of which\\ is a basic extra parameter.\end{tabular} \\ \hline
	\end{tabular}}}
\end{table}

\section{Our Hybrid Approach (\tool)}\label{sec:approach}
\tool takes an \emph{apk} as input and outputs the visualized storyboard (${S}$) with rich features for the app. 
Fig.~\ref{fig:workflow} shows the overview of our hybrid approach (named \tool): \revise{(1) First of all, \tool instruments the \emph{apk} so that activities can be launched by third-parties. (2) \textit{Static extraction} includes \textit{ATG extraction}, which leverages static program analysis to obtain relatively complete ATG. Meanwhile, the required {ICC data} (i.e., Activity launching parameters shown in Table~\ref{tbl:iccdata}) can be extracted through control- and data-flow analysis (refer to Section \ref{subsubsec:icc}). (3) \textit{Dynamic UI page rendering} launches the activities registered in the app one by one with the extracted ICC parameters. Meanwhile, it can also augment ATG through \textit{dynamic UI component exploration}. After that, we can obtain a comprehensive ATG with rendered UI pages. (4) Moreover, the other rich features, such as layout code, Activity code, UI component, and call graphs are collected.} (5) \tool then visualizes the storyboard of the app with all the extracted features in a webpage.

\begin{figure}
	\centering
	\includegraphics[width=0.45\textwidth]{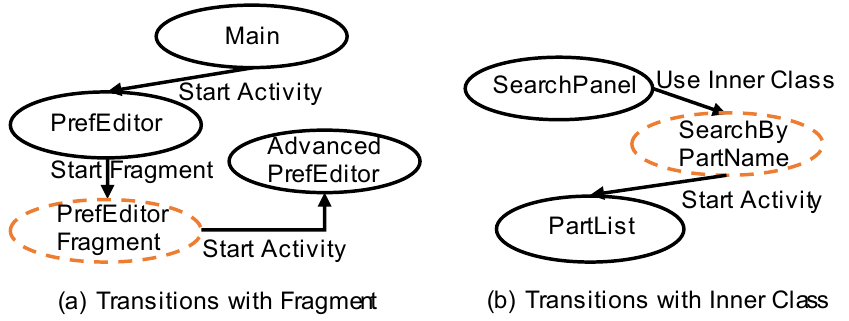}
	\caption{Activity transitions between activities and fragment, inner class}
	\label{fig:atg}
\end{figure}

\subsection{APK Instrumentation}\label{subsec:instrumentation}
\revise{In terms of the \textit{APK instrumentation}, we {first decompile the target apk and set} ``exported=true'' in the AndroidManifest.xml file for each activity to enable the launching process by third-parties. We then repackage it to a new installable APK file and sign it to ensure its usability. Note that the repackaged apps are only used for the experimental purpose, and all the experiments are conducted in a controlled environment. The repackaged apps will not be released for commercial use.}

\subsection{Static Extraction}\label{subsec:static}
\revise{Static extraction mainly contains two steps: ATG extraction and ICC data extraction for dynamic UI page rendering in the next phase.}

\subsubsection{{ATG extraction}}
\revise{Activity transition in fragment and inner class are representative and widely-used components in real-world apps. According to our study on 150 randomly selected real apps (75 open-source and 75 closed-source apps used in RQ1 (\S~\ref{subsec:rq1})), we find 44 apps use Activity transitions in fragment and 84 apps use Activity transitions in inner class.}
Before extracting activity transitions in inner classes and fragments, we illustrate the transitions in them. Fig.~\ref{fig:atg} (a) is the sub ATG of \emph{Vespucci}~\cite{vespucci}, a map editor. Firstly, activity {Main} starts PrefEditor, in which PrefEditorFragment is started. And PrefEditorFragment further starts AdvancedPrefEditor. Specifically, as shown in Listing~\ref{fig:fragment}, fragments can be added to an activity in two ways: (1) by invoking fragment modification API calls, e.g., ``replace()'', ``add()'', and further leveraging ``FragmentTransaction.commit()'' (lines 3-4) to start the fragment; (2) By using ``setAdapter'' (line 7) to display the fragment in a certain view (e.g., ViewPager). The started PrefEditorFragment then starts a new activity (i.e., AdvancedPrefEditor). Fig.~\ref{fig:atg} (b) shows the sub ATG of ADSdroid, where SearchPanel uses an inner class SearchByPartName to handle time-consuming operations as shown in Listing~\ref{fig:inner}. After finishing the task, it starts an activity PartList by invoking ``StartActivity()'' (line 4). In this example, our goal is to extract activity transitions: Main$\rightarrow$PreEditor, PreEditor$\rightarrow$AdvancedPrefEditor, and SearchPanel$\rightarrow$PartList. 

Algorithm~\ref{algo:static} details the extraction of ATG. Specifically, it takes as input an ${apk}$, and outputs the activity transition graph (${atg}$).
We first initialize $atg$ as an empty set (line 1), which stores the activity transitions gradually. We then generate the call graph ($cg$) of the given ${apk}$. For each method ($m$) in each class ($c$), if there exists an activity transition, we first get the target activity ($callee\_act$) by analyzing the data in \emph{Intent} \rTwo{via getTargetAct()} (lines 4-8). 
\rTwo{Specifically, for each explicit activity transition, the target activity is explicitly indicated in the Intent object where an intent variable usually either explicitly declares the callee activity or uses a variable defined before or other types of implementation to indicate the target activity. We first analyze which intent constructor it creates (Intent has various constructors to receive different kinds/numbers of parameters), and then track the parameter that indicates the target activity by data-flow analysis. Finally, we can obtain the target activity ($callee\_act$).}
If the method ($m$) is in an inner class, we regard the outer class as the activity that starts the target activity and add the transition to $atg$ (lines 9-11). Take Fig.~\ref{fig:atg}~(b) as an example, we add an edge SearchPanel$\rightarrow$PartList to $atg$. 

\begin{lstlisting}[language=Java,caption={Simplified code snippet of Fragment},label={fig:fragment}]
public class PrefEditor{... //Using replace/add
	PrefEditorFragment pref = new PrefEditorFragment();
	FragmentTransaction.replace(R.id.content,pref);
	FragmentTransaction.commit(); 
}
public class PrefEditor{... // Using setAdapter
	ViewPager.setAdapter(getSupFragmentManager(), new PrefEditorFragment());
}
\end{lstlisting}

\begin{lstlisting}[language=Java,caption={Simplified code snippet of Inner Class},label={fig:inner}]
public class SearchPanel{...
	private class SearchByPartName extends Asynctack<>{...
	    Intent intent = new Intent(MainActivity.this,PartList.class);
	    startActivity(intent); 
	} 
}
\end{lstlisting}

If $m$ is in a fragment, we construct the relation between the fragment ($caller\_frag$) and the target component (lines 12-13). 
\rTwo{Specifically, for each activity transition, we first locate the class $c$ that starts a new activity according to $m$, and then check the super class of it. If it extends a fragment, we then set $caller\_frag = c$. 
In fact, in terms of extracting the target activities from explicit transitions, there is no difference between extracting activities started by activity and fragment.}
Note that this relation \rTwo{between the caller fragment and the target activity} does not represent the actual component transition, we optimize it by identifying the activities that start the fragment in lines 18-21. 
\rTwo{Specifically, to identify the activities that bind a specific fragment, we investigate different types of methods that bind activities and the corresponding fragments, 
where fragments are operated (e.g., removed, added, replaced, and setAdapter) using specific APIs, and we can track specific APIs 
to identify the activity corresponding to a specific fragment.}
After that we update $atg$ by merging fragment relations to construct the actual activity transitions (line 22). For example, as shown in Fig.~\ref{fig:atg}~(a), we first obtain the relations PrefEditorFragment$\rightarrow$AdvancedPrefEditor, PrefEditor$\rightarrow$PrefEditorFragment, then we merge it to PrefEditor$\rightarrow$AdvancedPrefEditor to represent the actual activity transition. For method $m$ that is neither in an inner class nor a fragment, we backward traverse $cg$ starting from $m$ to obtain all the activities that start the target activity ($callee\_act$), then add them to $atg$ (lines 14-17).

\begin{algorithm2e}[t]
	\setcounter{AlgoLine}{0}
	\caption{Static ATG Extraction}
	\label{algo:static}
	\DontPrintSemicolon
	\SetCommentSty{mycommfont}
	\KwIn{$apk$}
	\KwOut{$atg$: Activity transition graph, including Activity and Service}
	$atg$ $\leftarrow \emptyset$ \;
	$cg$ $\leftarrow$ getCallGraph($apk$) \;
	$all\_classes$ $\gets$ getAllClasses($apk$) \;
	\ForEach{$c$ $\in$ $all\_classes$}{
		$methods$ $\gets$ getClassMethods($c$) \;
		\ForEach{$m$ $\in$ $methods$}{
			\If{hasActivityTransition($m$)}{
				$callee\_act$ $\gets$ getTargetAct($m$)\;
				\If{isInnerClass($c$)}{
					$caller\_act$ $\gets$ outerClass($c$) \;
					$atg$.addPair($caller\_act$, $callee\_act$)\;
				}
				\ElseIf{isInFragment($m$)}{
					$atg$.addPair($caller\_frag$, $callee\_act$)\;
				}
				\Else{
					$caller\_acts$ $\gets$  getCallerAct($m$, $cg$)\;
					\ForEach{$act$ $\in$ $caller\_acts$}{
						$act$.addPair($act$, $callee\_act$)\;
					}
				}	
			}
			\tcp*[h]{Optimize $atg$}\;
			\If{startFragment($m$)}{
				$caller\_acts$ $\gets$  getCallerAct($caller\_frag$)\;
				\ForEach{$act$ $\in$ $caller\_acts$}{
					$atg$.addPair($act$, $callee\_frag$)\;
				}
				updateATGIfNeeded($atg$)
			}	
		}
	}
	\Return{$atg$}
\end{algorithm2e}

\SetKwInput{KwInput}{Input}
\SetKwInput{KwInput}{Input}
\SetKw{Let}{let}
\SetKw{Continue}{continue}

\subsubsection{{ICC data extraction}}\label{subsubsec:icc}
As aforementioned in \S~\ref{sec:background-icc}, to successfully launch an activity, data that are required to render the target UI page should be provided, including the \emph{primitive attributes} and \emph{extra parameters} listed in Table~\ref{tbl:iccdata}. Algorithm~\ref{algo:iccdata} details the extraction process of ICC data. 
\rThree{We highlight that the data-flow analysis for ICC data extraction is one of the core phases in \tool, which obviously improves the ability of UI page rendering (c.f. \S~4.3). 
}

As shown in Algorithm~\ref{algo:iccdata}, it takes an \textit{apk} as input, and outputs the ICC data required to launch each activity. Specifically, we first obtain the call graph, all class instances, and the AndroidManifest.xml file by decompiling the apk, and the output $icc\_data$ is initialized as an empty set (Lines 1-4). We then traverse the classes to identify activities. For each activity $act$, we use the function \texttt{getParamters()} to obtain the required parameters (including \emph{primitive attributes} and \emph{extra parameters}) for launching the activity (Lines 9-31). As for the \textbf{primitive attributes}, we obtain them (if any) from the manifest file by parsing the corresponding fields, such as ``action'' and ``category'', and then save it in $para$ (Lines 11-12). Sometimes primitive attributes are also declared in the source code, and the extraction method is similar to that of extra parameters.

\begin{algorithm2e}[t]
	\setcounter{AlgoLine}{0}
	\caption{ICC Data Extraction}
	\label{algo:iccdata}
	\DontPrintSemicolon
	\SetCommentSty{mycommfont}
	\KwIn{$apk$}
	\KwOut{$icc\_data$ <$act$, $para$>: ICC data of each activity for Activity launching}
	$icc\_data$ $\leftarrow \emptyset$ \;
	$cg$ $\leftarrow$ getCallGraph($apk$) \;
	$all\_classes$ $\gets$ getAllClasses($apk$) \;
	$mani \gets$ getManifest($apk$)\;
	\ForEach{$c$ $\in$ $all\_classes$}{
		\If{isActivity($c$)}{
			\tcp*[h]{Get parameters for activity $c$}\;
			$para \gets$ getParameters($c$, $cg$, $mani$) \; 
			$icc\_data$ = $icc\_data$ $\bigcup$ <$c$, $para$>\;
		}
	}
	\SetKwFunction{FMain}{getParameters}
	\SetKwProg{Fn}{Function}{:}{}
	\Fn{\FMain{$act$, $cg$, $mani$}}{
		$para$ $\gets \emptyset$ \;
		\tcp*[h]{Get primitive attributes from manifest}\;
		$attr, value \gets$ getPrimitiveAttr($c$, $mani$)\;
		$para \gets para \bigcup$ <$attr$, $value$> \;
		\tcp*[h]{Get extra parameters from source code}\;
		$methods_{lc}$ $\gets$ getLifecycleCallbacks($act$) \;
		\ForEach{$m$ $\in$ $methods_{lc}$}{
			$type$, $key$ $\gets null$
			\; 
			$para \gets$ getExtras($m$, $para$);
		}
		\KwRet $para$\;
	}
	
	\SetKwFunction{FMain}{getExtras}
	\SetKwProg{Fn}{Function}{:}{}
	\Fn{\FMain{$m$, $para$}}{
	\If{hasExtraParameters($m$)}{
			$extras \gets$ getAllExtras($m$)\;
			\ForEach{$e \in extras$ }{
				$key$ $\gets$ getKey($e$)\;
				$type$ $\gets$ getValueType($e$)\;
				$para \gets$ $para \bigcup$ <$key$, $type$> \;
			}
		}
	\Else{
				\revise{$m_{callee}$ $\gets$  getCalleeMethod($m$, $cg$)\;}
				\While{\revise{$m_{callee}\neq null$}}{
				   \revise{$para \gets$ getExtras($m_{callee}$, $para$)\;}
					\revise{$m_{callee}$} $\gets$ getCalleeMethod(\revise{$m_{callee}$}, $cg$)\;
				}
			}
	\KwRet $para$\;
	}
	
	\Return $icc\_data$
\end{algorithm2e}

As for the \textbf{extra parameter} extraction, we first identify methods related to activity lifecycle (denoted by $methods_{lc}$), such as \textit{onCreate}() and \textit{onStart}() since extra parameters in these methods are related to page rendering. For each lifecycle callback (i.e., method) $m$, if it invokes specific APIs (e.g., \textit{getStringExtra}, \textit{getBundle}) to get the ICC extra data from the previous activity, we obtain the key through backward data-flow analysis and the value type of each extra parameter based on the corresponding APIs. We then save them in $para$ (Lines 19-24).
\rTwo{Specifically, as for the \textit{key}, whose main purpose is to get the attached data transferred from the source activity to the target activity, therefore, it is usually presented using constant strings and can be directly extracted from the code according to the specific APIs. As for the \textit{value type}, we can get it according to the specific APIs of value types such as getStringExtra and getBooleanExtra. For example, btd =getIntent().getStringExtra(``returnKey1"), the key we obtained is ``returnKey1'', the value type is ``String'', and the ICC data is saved as $<$returnKey1, String$>$, we will provide a string value for the key returnKey1 at runtime to launch the target activity.
}
In some cases, the extra parameters are not directly declared in the lifecycle methods, but in the methods that the lifecycle methods invoke, which would also affect the UI rendering if not provided with proper parameters. To tackle this situation, we first obtain the methods that the lifecycle callbacks invoke according to the call graph $cg$ (Line \revise{26}), and iteratively explore each method to obtain the potential extra parameters with their key and value types by invoking $getExtras$ method (Lines \revise{28-29}). After obtaining the parameters for activity $act$, we store it together its required parameters to $icc\_data$ for further UI page rendering and exploration.

\subsection{Dynamic UI Page Rendering}\label{subsec:dynamic}
\revise{Dynamic UI page rendering mainly contains two steps: \textit{UI page rendering}, which {launches each activity} dynamically based on the extracted ICC data; \textit{UI component exploration}, which {augments static ATG by exploring all interactive components of each activity to identify more activity transitions together with UI pages.}} \rThree{Note that the static UI page rendering method used in \oldtool by leveraging layout code transformation may lead to a big visual difference between the rendered pages and the real pages like Fig.~9 (a), the reasons are explained in \S~5.2.2. However, there are no such limitations in StoryDistiller which uses the dynamic UI page rendering with the ICC data extracted by the data-flow analysis.}

\subsubsection{UI page rendering}\label{subsec:rendering}
After generating the activity transitions between different pages, we now aim to render the corresponding UI pages by exploring each activity of the app. Our goal is to render/explore as many UI pages as possible to visualize the transitions between activities. To the best of our knowledge, basically, there exist two methods rendering/exploring the UI pages: (1) Dynamic app testing tools such as {Monkey}~\cite{monkey} and {Stoat}~\cite{su2017guided}, which aim to explore as many UI pages as possible by dynamically running the apps to detect more bugs, however they are demonstrated to only achieve \textasciitilde35\% activity coverage, which is far away from representing the complete relations between activities. (2) Static UI page rendering. Chen et al.~\cite{chen2019storydroid} proposed to render UI pages by first converting dynamic/hybrid layout to static layout since they found 62.3\% apps construct their UI pages by adopting dynamic/hybrid layout, and developing a dummy app to launch each activity with the help of static layouts. However, the rendering largely relies on the layout conversion process, causing incomplete or error rendering of the UI pages if the conversion process is incomplete.

\revise{To this end, we propose to render and launch activities dynamically with the help of the extracted ICC data and Android toolkit, and take screenshots accordingly.} This approach has several advantages over the existing methods: (1) It does not need to generate test cases to run the app like dynamic app testing, but directly launch all activities one by one, which addresses the limitation that the test cases may not reach all the activities successfully; (2) It considers the data transferred from the previous activity that is essential for rendering the current activity, which alleviates the limitation of improper conversion process in \oldtool~\cite{chen2019storydroid}. The detail of the rendering process is as follows.

\revise{For each activity, if it requires parameters to launch, we provide it with a random dummy value according to its required data types (e.g., String/Integer/Boolean). 
\rTwo{As for the dummy value, we extracted the data defined the layout files when exploring apps and randomly choose values from them for different data types.}
In this way, we can append all parameters needed for activity launching. 
\rTwo{For example, if the extracted parameters of one Activity are <``userid'', Integer> and <``username'', String>. We will use the command: ``\texttt{adb shell am start -n pkg/pkg.activityname --ei userid 2 --es username Alice}'' to launch the current Activity, where \texttt{--ei} and \texttt{--es} refer to the data types of the parameters are Integer and String, respectively. More required extra parameters can be appended. For other data types, there are also corresponding commands, such as \texttt{--ez} for Boolean and \texttt{--ef} for float.}
Besides, to eliminate side-effect between different activities during launching, we provide a fresh state for each activity by forcing stop the previous launched ones.} For activities that fail to launch due to app crashes or permissions required, we dump the layout hierarchy of the current activity and analyze it to check whether it contains keywords (e.g., ``has stopped'' and ``keeps stopping'' for app crashes, ``ALLOW'' and ``DENY'' for permission requests). When the app crashes, we stop the app and set it to the original state (i.e., a fresh state for another activity to launch). When the app requests permission from users, we automatically grant it to make it render the UI page normally.

Note that, the activity that is actually launched may be different to what is intended to be launched. For example, we intend to launch an activity called ``{NewsDetailActivity}'', however this activity requires user credentials (e.g., user name and password). Without valid user credentials, it would jump to the ``sign in'' or ``sign up'' page. Thus the actual launched activity would be the ``signInActivity''. Considering such situations, to avoid assigning incorrect activity names to the launched UI pages, we obtain the current launched activity by retrieving the top activity from the back stack through the Android running system. This strategy also addresses the code obfuscation problem on activity names, which is better than the solution proposed in \oldtool~\cite{chen2019storydroid}, i.e., inferring semantic names based on the layout tree similarity.

\subsubsection{UI component exploration}
\revise{Although the completeness of ATG is much better than the existing static method such as IC3 and dynamic method such as Stoat according to the comparison experiments in StoryDroid~\cite{chen2019storydroid}, some of the {important} activity transitions are still missing due to the limitation of the pure static method. In this paper, we propose to explore interactive components on each page and augment ATG. Specifically, when the UI page is rendered successful (\S~\ref{subsec:rendering}), \tool follows two steps to conduct dynamic UI component exploration.}

\revise{
Firstly, we parse the layout code of each rendered activity and extract each \textit{interactive component} (e.g.,  \textit{ImageButton}, \textit{Button}, and \textit{clickable TextView}) together with its attributes, including UI component id, component description, etc. Secondly, we trigger each interactive component on the rendered activity by using UIAutomator\cite{uiautomator}. If the behavior triggers the launching of another activity and the transition is not included in the current ATG, we add the new explored transition pair into the ATG.} By leveraging the hybrid ATG construction approach, we are able to obtain a more complete ATG for the demonstration of storyboards.

\subsection{Rich Feature Extraction and Implementation}
\subsubsection{Feature Extraction}
To visualize the storyboards of Android apps with all rich features, we highlight the extracted rich features for different software engineering tasks. Specifically, as shown in Fig.~\ref{fig:workflow}, we extract {8} kinds of features, including ATG, UI page, activity name, layout code, activity code, call graph, and UI components with their attributes. Among them, ATG, UI page, activity name, and call graph are extracted in \S~\ref{subsec:static}-\S~\ref{subsec:dynamic} to achieve specific tasks. For activity code, we extracted the corresponding code by decompiling the APK file using the reverse-engineering tool. For layout code, we obtain them when rendering UI pages by dumping the layout for the current activity, which is the actual layout for the launched activities. For UI components and their attributes, we first identify the boundary and the attributes of each component (e.g., ``Button'' and ``EditText'') from the layout code, and crop each component according to the boundary.

\begin{figure}
	\center
	\includegraphics[width=0.5\textwidth]{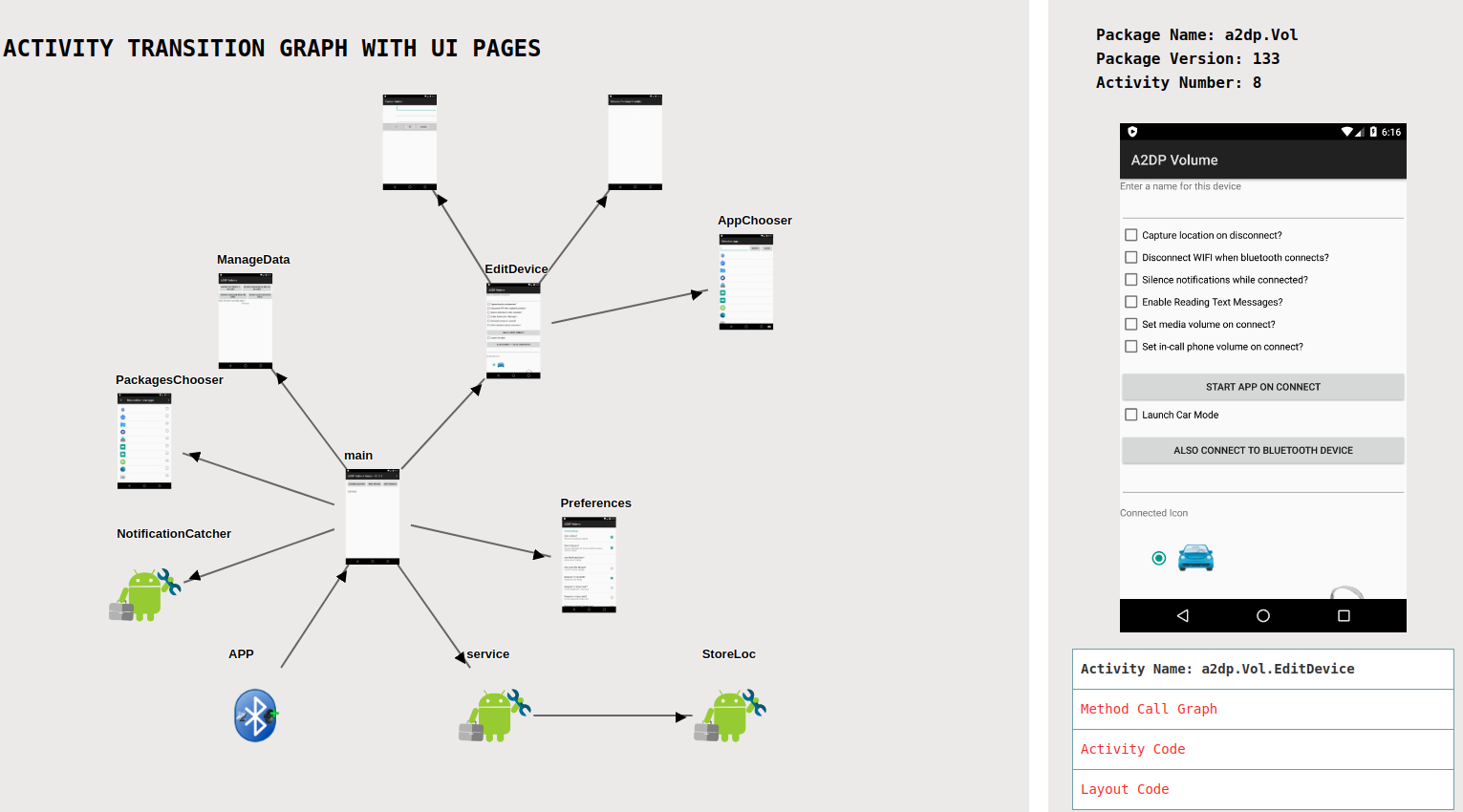}
	\caption{Web service of \tool}
	\label{fig:service}
\end{figure}

\subsubsection{Implementation}
We implement \tool as an automated tool, which is written in 4K lines of Java code, and 3K lines of Python code. \tool is built on top of several off-the-shelf tools: IC3, jadx~\cite{jadx} and Soot~\cite{soot}. We extend the Soot framework to extract inputs of UI page rendering, such as ATG and ICC data, and get the call graphs from apks. Activity transition extraction is built on IC3 to obtain a comparatively complete ATG. jadx is used to decompile the apk to obtain the source code for Android apps. ApkTool (v2.4.1)~\cite{apktool} is used to repackaged the apk to implement the instrumentation. 
\rTwo{We dump the actual activity names for each UI page from the console through activity back stack \cite{component_stack}. For the few cases where activity names lack semantics and users have demand to obtain the inferred activity name, the method proposed in StoryDroid \cite{storydroid} can also be applied.}
The used Android emulator (Nexus 5X) is running on Genymotion (v3.0.0) with Android 8.0, 4G RAM, and 1920$*$1080 resolution ratio. We use data-driven document (D3)~\cite{d3} to visualize \tool's results, which provides a visualized technique based on data in HTML, JavaScript, and CSS. As shown in Fig.~\ref{fig:service}, the visualization~\cite{chen2019storydroid} contains 4 parts: (1) ATG with activity names and corresponding UI pages; (2) The layout code of each UI page; (3) The functional code of each activity; (4) The components of each UI page with corresponding attributes, such as label and size; (5) The method call relations within each activity.

\section{Evaluation of \tool} \label{sec:effectiveness_eval}
In this section, we evaluate the effectiveness and the usefulness of \tool based on the following three research questions: 

\noindent {\bf RQ1:} \revise{Can \tool extract a more complete ATG in terms of more transitions and higher activity coverage compared with existing ATG exploration tools (i.e., {IC3}~\cite{octeau2015composite}, Gator~\cite{gator}, Stoat~\cite{su2017guided}, and StoryDroid~\cite{chen2019storydroid})?}

\noindent {\bf RQ2:} \revise{Can \tool render more UI pages with higher UI similarity compared with StoryDroid?}

\noindent {\bf RQ3:} \revise{Can \tool help explore and review the functionalities of Android apps effectively and efficiently?}

\subsection{\revise{RQ1: Effectiveness of Hybrid ATG Extraction}}\label{subsec:rq1}

\subsubsection{Setup} \revise{To investigate the capability of constructing ATG, we randomly download 75 apps from Google Play Store (closed-source apps) and 75 apps from F-Droid~\cite{fdroid} (open-source apps) as subjects to demonstrate the effectiveness of ATG extraction on real-world apps. We compare \tool with four existing ATG exploration tools including three static methods (i.e., IC3~\cite{octeau2013effective}, Gator~\cite{gator}, and StoryDroid~\cite{chen2019storydroid}), and one dynamic method, i.e., Stoat~\cite{su2017guided} which has been demonstrated to be more effective on app exploration than other tools such as {Monkey}~\cite{monkey}. For some closed-source apps, IC3 and Gator take more than one hour to extract ATG probably due to some internal errors, therefore, we set a timeout of 30 minutes for each app which is sufficient to explore most of the apps. For Stoat, we run each app for 30 minutes. As for the evaluation metrics, we use the number of \textit{activity transition pair} and \textit{activity coverage} to demonstrate the performance of each tool. ``activity coverage'' is computed as the number of unique activities in the ATG over the total number of activities declared in the app.}

\subsubsection{Results of RQ1}

\begin{figure}
	\center
	\includegraphics[width=0.45\textwidth]{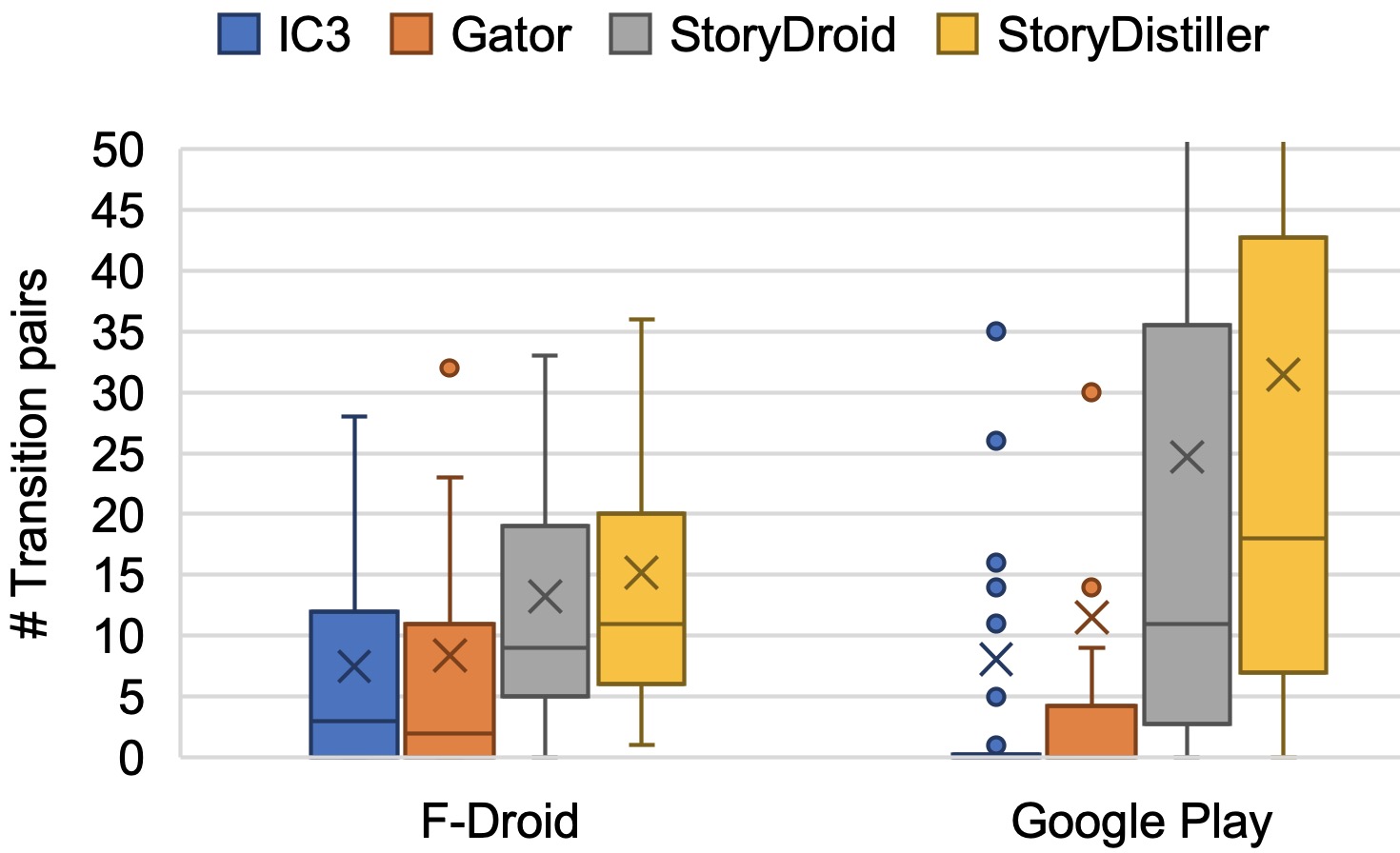}
	\caption{Comparison of transition pairs}
	\label{fig:pairs_cmp}
\end{figure}

\noindent \textbf{Activity transition pairs.}
\revise{Fig.~\ref{fig:pairs_cmp} shows the result of tool ability in terms of extracting activity transition pairs. It can be seen that \tool outperforms the other three static tools for both open-source apps and closed-source apps. More specifically, \tool is able to extract 15.2 and 31.4 transition pairs on average for each open-source app and each closed-source app, receptively. Compared with \oldtool, \tool improves over 20\% transition pairs, which benefits from the proposed dynamic UI component exploration. Compared with IC3 and Gator, \tool increases more than twofold (7.78 for IC3, 9.96 for Gator vs. \textbf{23.30}) on all these selected apps.}

\revise{As for \tool, it can extract activity transitions with respect to all the features such as fragments, inner classes, and callbacks. Since we extract transitions by using particular APIs (e.g., {StartActivity}, {StartActivityForResult}, and {StartActivityIfNeeded}) that start new activities by leveraging data-flow analysis, the extracted transitions are more accurate. 
\revise{To investigate the contribution of fragment and inner class to ATG, we record the number of apps that use fragment or inner class to start new activities. The result shows 44 apps use fragments and 84 apps use inner class to start new activities, indicating the popularity of using these two types to build activity transitions in real scenarios.}
Besides, the dynamic UI component exploration can also augment ATG. Even though, \tool sometimes may still miss some transitions due to the limitation of the underlying tools such as decompilation failures or extraction errors of certain classes.
Besides, developers may self-define some methods to start new activities instead of using the default patterns (\textit{c.geo}~\cite{geo} open-source app), causing some activity transition pairs cannot be identified and extracted. Similarly, intent overloading~\cite{overloading} would also lead to missing activity transitions (\textit{FBReader: Favorite Book Reader}~\cite{fbreader}). To some extend, dynamic UI component alleviates this problem and augment the transitions effectively. Overall, the results shown in Fig.~\ref{fig:pairs_cmp} demonstrate the effectiveness of \tool on extracting activity transitions over other existing tools.}

\revise{Compared with IC3, \tool has advantages on inner classes, fragments, and callbacks when extracting activity transitions, which has been evaluated in \oldtool~\cite{chen2019storydroid}. However, according to our investigation, for intent overloading with complex parameters, IC3 can extract partial activity transitions statically. Therefore, to obtain a comparatively complete ATG and maximize the activity coverage, we implemented \tool by integrating the transition results of IC3.}

\begin{figure}
	\center
	\includegraphics[width=0.45\textwidth]{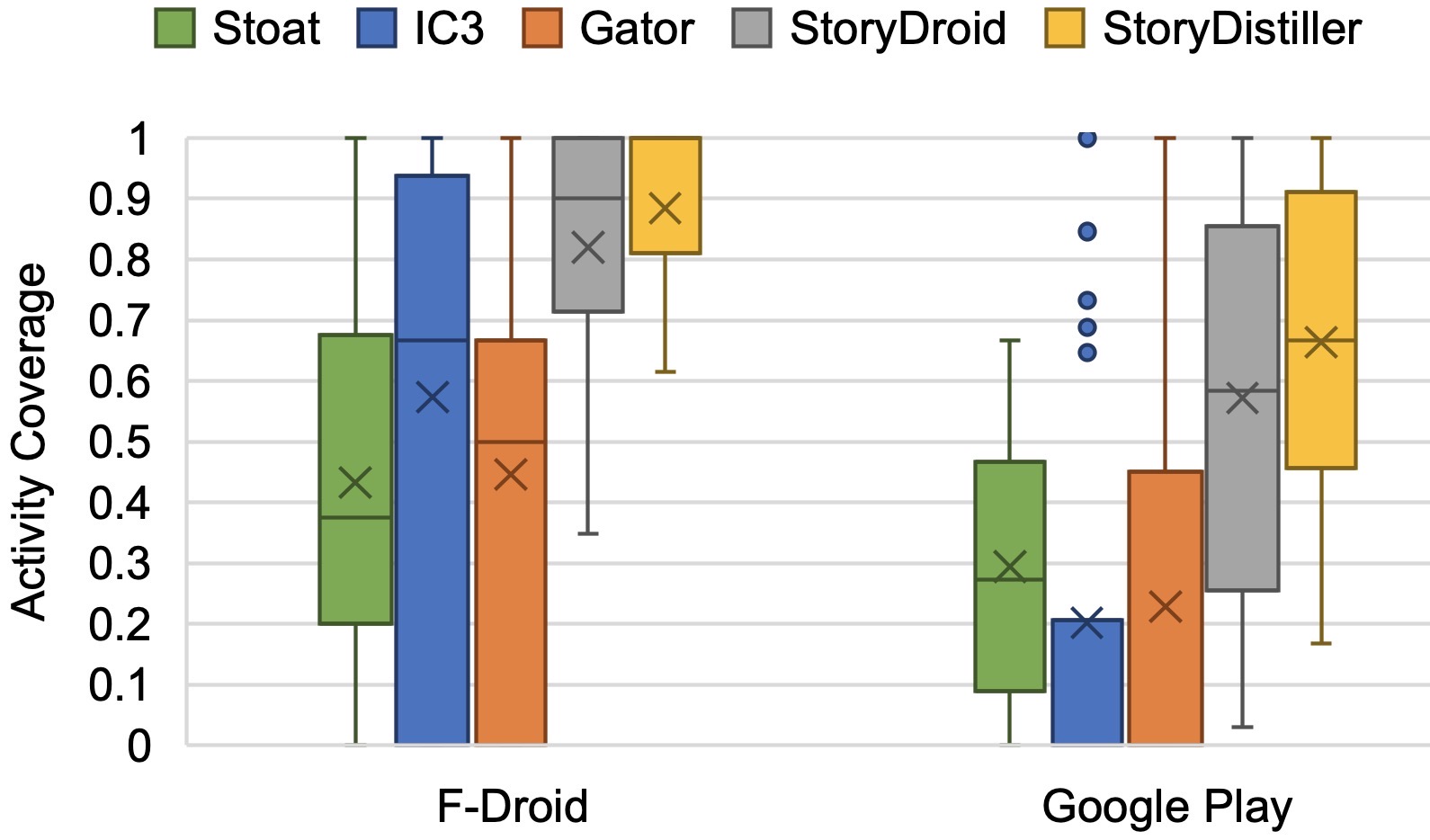}
	\caption{Comparison of activity coverage}
	\label{fig:coverage_cmp}
\end{figure}

\noindent \textbf{Activity coverage.}
Fig.~\ref{fig:coverage_cmp} depicts the activity coverage results of each tool. Compared with the dynamic method, on average, \revise{\tool outperforms Stoat in terms of activity coverage, achieving 88.5\% (vs. 43.2\%) and 66.4\% (vs. 29.4\%) coverage on open-source apps and closed-source apps, respectively. In addition, \tool costs much less time (i.e., 8.50 minutes on average) to extract and render the activities than Stoat (i.e., 30 minutes). The time cost includes the apk instrumentation and UI page rendering. As for the comparison results with static methods, the performance trend is similar to that of activity transition pairs. \tool still outperforms other tools, achieving nearly 80\% coverage on average. Compared with \oldtool, \tool improves over 10\% activity coverage.}

\tool does not cover all the activities for some apps due to the following reasons: (1) the limitation of reverse engineering techniques, some classes and methods cannot be decompiled from apks, causing failures in extraction of activity transition and coverage. That situation is more severe in closed source apps due to packing~\cite{packers} and code obfuscation techniques~\cite{proguard,dasho}. (2) Another reason is the dead activities (no transitions), such as unused legacy code and testing code in apps. We also investigate the reasons why dynamic exploration tools such as {Stoat} achieve low activity coverage: (1) Login requirement. For example, {Stoat} fails to explore {Santander} which is a banking app requiring login using password or fingerprint. (2) Lack of specific events. For example, {Open Training} is a fitness-training app, which can create fitness plans by swiping across the screen. However, {Stoat} does not support such events, resulting in low coverage.


\begin{tcolorbox}[size=title,opacityfill=0.1,breakable]
\revise{\textbf{Answer to RQ1.} \tool outperforms the static methods (e.g., IC3, Gator, and \oldtool) in terms of activity transition pairs (\textbf{23.3} vs. 7.8 in IC3, 10.0 in Gator, and 18.2 in \oldtool), and the dynamic method (e.g., Stoat) in terms of activity coverage (\textbf{77.5\%} vs. 36.3\% in Stoat). Therefore, \tool is able to obtain a more complete activity transition graph compared with existing tools.}
\end{tcolorbox}

\subsection{RQ2: Effectiveness of UI Page Rendering}\label{subsec:rq2}
\subsubsection{Setup}
To investigate the effectiveness of UI page rendering, \revise{we compare \tool (dynamic method) and \oldtool (static method) in terms of the ratio of rendered pages and the UI similarity of rendered pages, by using the 150 Android apps in RQ1.} Specifically, (1) we first investigate the ratio of UI pages (activities) that are successfully launched in each app, denoted by $LaunchR$. 
\[
LaunchR_i=\frac{N_{i}^{Launched\_act}}{N_{i}^{All\_act}} \times 100\%
\] 
Where $N_{i}^{All\_act}$ indicates the number of activities declared in the AndroidManifest.xml file in the $i^{th}$ app. (2) We further investigate whether the functionalities of the launched pages are clearly displayed, i.e., users can easily and clearly understand the functionality through the UI pages. 
\revise{To do it, we compute the visual similarity between the rendered UI pages and the real UI pages to demonstrate the rendering ability of \tool in practice. Note that the real UI pages are obtained by Monkey~\cite{monkey}.}

\begin{figure}
	\center
	\includegraphics[width=0.315\textwidth]{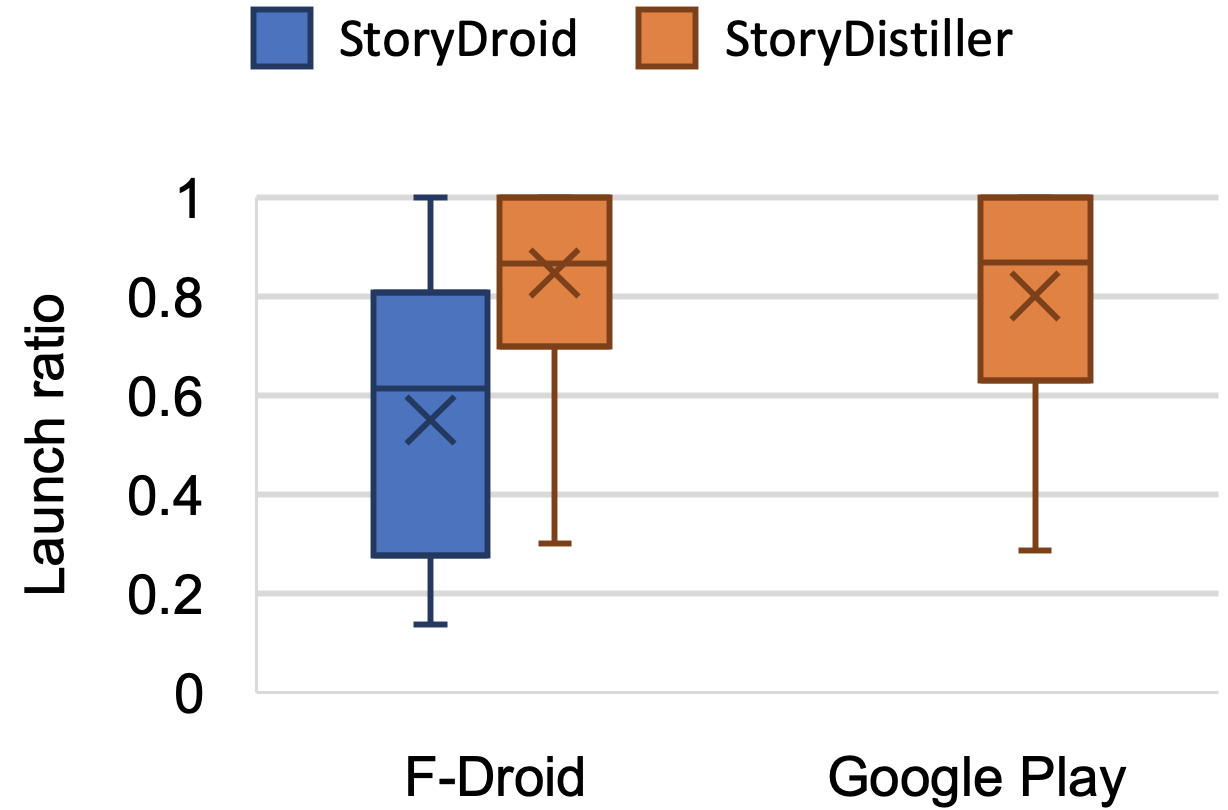}
	\caption{Comparison the Launch ratio of activities between \oldtool and \tool}
	\label{fig:cmp-ratio}
\end{figure}

\begin{figure*}
	\center
	\includegraphics[width=0.95\textwidth]{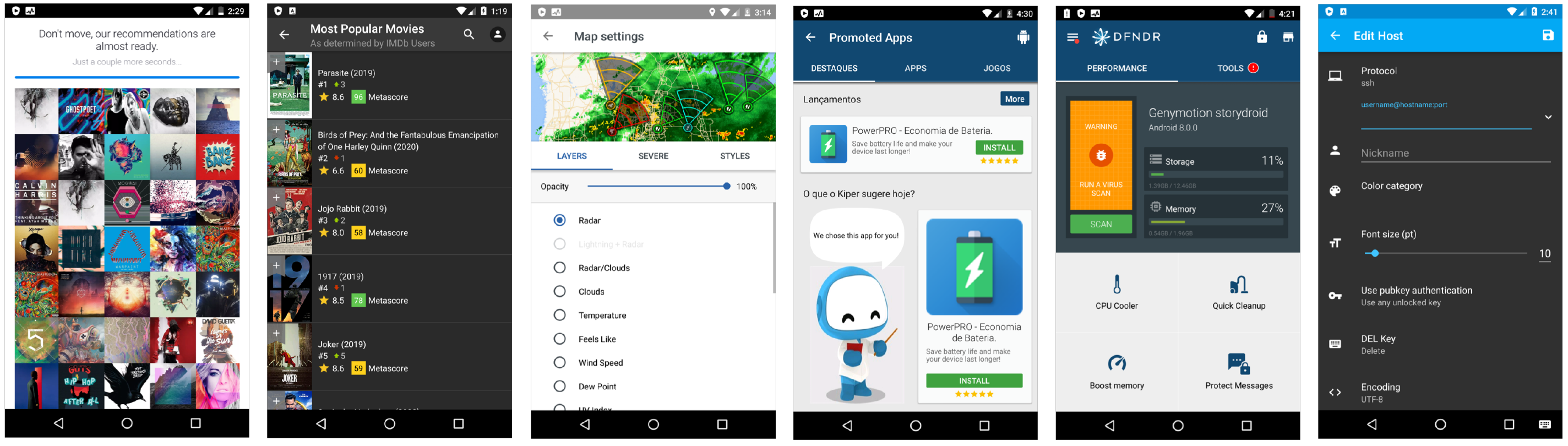}
	\caption{\revised{Examples of successfully rendered UI pages with diverse components}}
	\label{fig:good_cases}
\end{figure*}

\subsubsection{Results of RQ2}\label{subsub:RQ2}
\noindent \textbf{Launch ratio.} Fig.~\ref{fig:cmp-ratio} shows the distribution of the launch ratio of each app. \revise{Note that since \oldtool fails to launch activities due to the apk compilation failures for most of the closed-source apps (only 6 out of 75 closed-source apps), to avoid introducing bias based on such a small dataset, we only show its result on open-source apps in the box plot. The reasons for such a low launch ratio on closed-source apps are also described in this section.} We can see that on average, over 80\% activities (i.e., \rTwo{82.37\% for open-source apps, 80.14\% for closed-source apps shown in Table~\ref{tbl:ratio}}) can be launched successfully by \tool in our dataset, the remaining ones encounter crashes when being launched, usually caused by ``NullPointerException'' or ``ClassNotFoundException''. 
\revise{Besides, we further investigate the contribution of ICC data to activity launching, with the help of the extracted ICC data by \tool, there are 37.69\% additional activities for closed-source apps (29.48\% for open-source apps) being launched successfully.}

\revise{\rTwo{While \oldtool only launches 55\% activities for open-source apps, as shown in Table~\ref{tbl:ratio}}.} Fig.~\ref{fig:cmp-ratio} also indicates that apps from Google Play are more likely to get lower launch ratio, i.e., more cases at the bottom. \rTwo{Almost all the launch ratio of open-source (\rThree{i.e., 73 apps}) are over 50\%, and the lowest launch ratio is 30\% in our dataset shown in Table~\ref{tbl:ratio}}. 
However, as for closed-source apps, there are some cases (\rThree{i.e., 9 apps}) whose launch ratio are below 50\%. \rTwo{Specifically, the lowest rate achieves 28.27\% launch ratio shown in Table~\ref{tbl:ratio}.} \revise{A possible reason may be the more complex functionalities in closed-source apps, which is evidenced by the number of transition pairs in Fig.~\ref{fig:cmp-ratio} (a).} 


\begin{table}[t]\footnotesize
\centering
\caption{\rTwo{Comparison the Launch ratio of activities including average, minimum, and maximum ratio between StoryDroid and StoryDistiller}}
\rTwo{\begin{tabular}{c|c|c|c|c}
\hline
\multicolumn{2}{c|}{\textbf{Method}}  & \begin{tabular}[c]{@{}c@{}}Static\\ (StoryDroid)\end{tabular}  & \multicolumn{2}{c}{\begin{tabular}[c]{@{}c@{}}Dynamic\\ (StoryDistiller)\end{tabular}} \\ \hline
\multicolumn{2}{c|}{\textbf{Sources}} & F-Droid & F-Droid     & Google Play    \\ \hline
\multirow{3}{*}{\begin{tabular}[c]{@{}c@{}}\textbf{Launch}\\ \textbf{Ratio}\end{tabular}} & Avg. & 55\% & 82.37\% & 80.14\% \\ \cline{2-5} 
            & Min.           & 13.64\% & 30\%        & 28.27\%        \\\cline{2-5} 
            & Max.           & 100\%   & 100\%       & 100\%          \\ \hline
\end{tabular}
}
	\label{tbl:ratio}
\end{table}

\revise{The main reason that \oldtool fails to launch activities for most of the closed-source apps is apk compilation failures listed below, which are all mitigated by \tool.} 
(1) \revise{\textit{Due to missing necessary configuration files}.}
\oldtool supports rendering UI pages for open-source apps because it requires to obtain the configuration file of the project (e.g., \emph{build.gradle}\rThree{\footnote{\rThree{A \emph{build.gradle} file will be generated when creating a new Android project through Android Studio. We take this file as the default \emph{build.gradle} when closed-source apps render the UI pages in \oldtool.}}}) which includes necessary library dependencies and other configurations, however, the configuration file only appears in the source code. It cannot be obtained even by decompiling the apk files. (2) \revise{\textit{Due to user-defined components~\cite{user_defined_component}, complex grammar representations, and resource file errors (e.g., XML layout files), etc.}} Even for open-source apps, it is still difficult for \oldtool to obtain user-defined components and complex grammar representations (e.g., Syntactic Sugar~\cite{syntactic_sugar}) by using the proposed static method, causing rendering failures in these UIs. \revise{Last but not least, errors caused by resource files sometimes occur when we build the dummy app.}
These limitations cause \oldtool ineffective in many apps in our dataset.

\vspace{1mm}
\noindent \textbf{Visual similarity.} \revise{We compare the visual similarity between the real pages and the rendered UI pages by \tool and \oldtool to demonstrate the quality improvement of rendered UI pages based on the 150 apps in RQ1. We obtain real pages by leveraging Google Monkey~\cite{monkey} to dynamically explore UI pages and take screenshots, and select the overlapping activities of real ones and rendered ones by their activity names. We use two widely-used similarity metrics~\cite{nguyen2015reverse,chen2019storydroid,chen2019guisquatting}: mean absolute error (MAE) and mean squared error (MSE) to measure the visual similarity.}

\begin{figure}
	\center
	\includegraphics[width=0.48\textwidth]{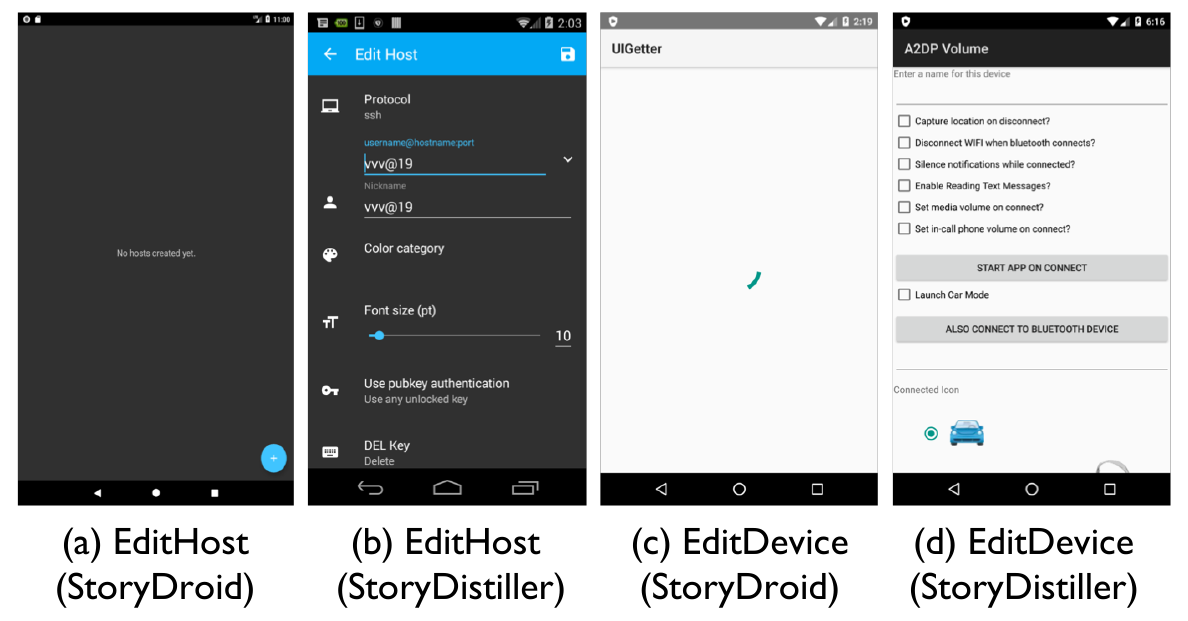}
	\caption{Examples of the same UI pages rendered by {\oldtool} and \tool}
	\label{fig:two_versions}
\end{figure}

\revise{The result shows that \oldtool only achieves about 80\% UI similarity, while \tool achieves 96.5\% and 91.6\% UI similarity in terms of MAE and MSE respectively.} Fig.~\ref{fig:good_cases} shows some real examples rendered by \tool, we can see that \tool can render UI pages with various types of components, such as RadioButton and ListView. Even for the UI pages using complex design structure or theme, multi-components, self-defined components, multi-images, or rich page color, \tool still performs well in most cases. Compared with \tool, \oldtool only uses testing data to replace real data for components such as ListView and GridView, which decreases the UI similarity compared with the real UIs. As shown in Fig.~\ref{fig:two_versions}, \oldtool cannot render such complex design structure or theme due to lack of data dependency, which would lose some main functionalities.  For example, Fig.~\ref{fig:two_versions} (a), rendered by \oldtool, shows that ``No hosts created yet'' without showing the main structure of the UI page due to lack of history data for the EditHostActivity. In contrast, Fig.~\ref{fig:two_versions} (b), rendered by \tool, displays all functionalities dynamically even for the save button (on the top right) with the real theme. Similarly, Fig.~\ref{fig:two_versions} (c) cannot be rendered like Fig.~\ref{fig:two_versions} (d) to demonstrate the real functionalities.

As we investigated, sometimes errors still occur in the rendered UI pages by \tool due to data loss in practice. We summarized six types as follows and show some real cases in Fig.~\ref{fig:bad_cases}. (1) \textit{Remote server data}. Fig.~\ref{fig:bad_cases} (a) shows an activity named ``PaylistActivity'', however, the detailed pay list information is lost since they are stored in the remote server. (2) \textit{Local database data}. Fig.~\ref{fig:bad_cases} (b) shows the profile of a user without detailed data since the data should be loaded from the local SQLite database. (3) \textit{Unauthorized access to webpages}. As shown in Fig.~\ref{fig:bad_cases} (c), it is a WebView page shows the terms and condition, however, due to unauthorized access, the WebView page fails to load. (4) \textit{Hardware support.} Fig.~\ref{fig:bad_cases} (d) is an app relying on hardware support (i.e., NFC). However, we conduct the UI page rendering on an Android emulator without required hardware support. (5) \textit{Login authentication.} Fig.~\ref{fig:bad_cases} (e) failed to be rendered due to the login authentication. Only users with valid authentication information can get access to the page, as indicated by the activity name. (6) \textit{Long loading time.} Fig.~\ref{fig:bad_cases} (f) is a map app. Due to the inadequate rendering time, the map is not rendered completely.

\revised{Although some specific data is not loaded or rendered successfully, the rendered information together with the activity names are still enough for users to understand the functionality of these pages.
For instance, for the cases in Fig.~\ref{fig:bad_cases} (a) (b) (f), we still can know the core logic of the activities. For the other three cases (i.e., Fig.~\ref{fig:bad_cases} (c) (d) (e)), the activity names contain rich semantics, which can help users understand the core logic.}


\begin{tcolorbox}[size=title,opacityfill=0.1,breakable]
\revise{\textbf{Answer to RQ2.} \tool achieves \textasciitilde80\% launch ratio of activities for each app on average, which is much better than \oldtool with only \textasciitilde55\% launch ratio when rendering UI pages on our dataset. Moreover, the rendered UI pages by \tool achieve a high UI similarity compared with \oldtool (\textasciitilde80\% vs \textbf{\textasciitilde95\%}).}
\end{tcolorbox}

\begin{figure}
	\centering
	\begin{subfigure}[t]{0.23\textwidth}
		\centering
		\includegraphics[width=0.95\textwidth]{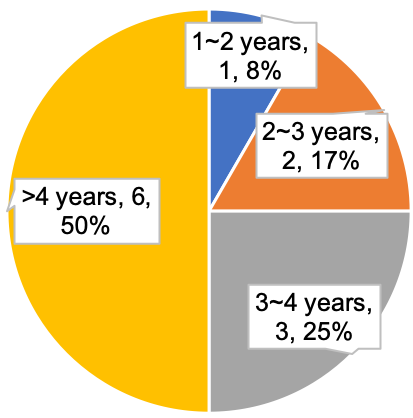}
		\caption{Years of Android device usage}
		\label{fig:usage}
	\end{subfigure}%
	\hfill
	\begin{subfigure}[t]{0.23\textwidth}
		\centering
		\includegraphics[width=0.95\textwidth]{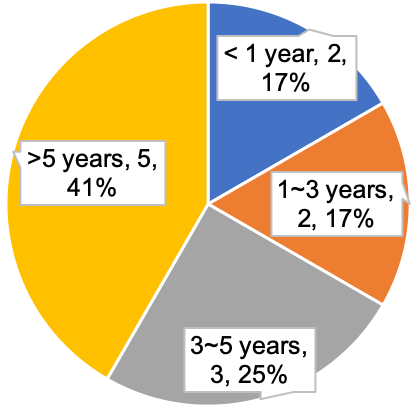}
		\caption{Years of conducting Android-related work}
		\label{fig:topic}
	\end{subfigure}
 	\caption{Distribution of participants}
 	\label{fig:distribution}
 \end{figure}
 
\begin{figure*}
	\center
	\includegraphics[width=0.95\textwidth]{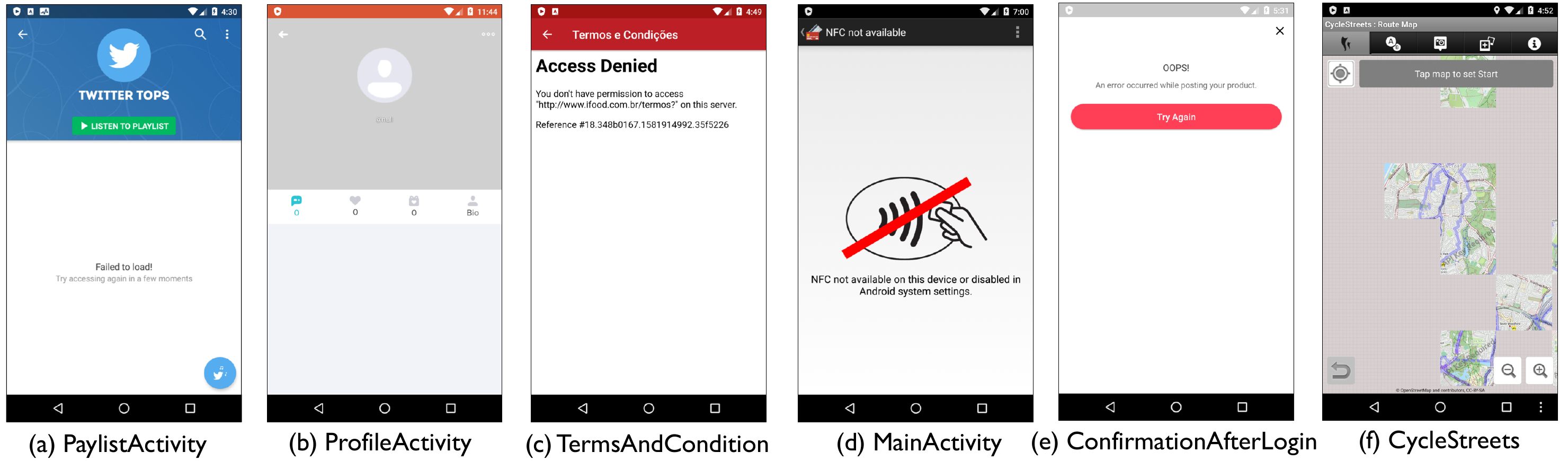}
	\caption{\revised{Examples of rendered pages with clear functionalities but with some data loss}
	}
	\label{fig:bad_cases}
\end{figure*}

\subsection{Usefulness Evaluation of \tool}\label{sec:usefulness_eval}
Apart from effectiveness evaluations, we further conduct a user study to demonstrate the usefulness of \tool. Our goals are to check whether \tool can help explore and \revised{review} the functionalities of apps effectively {and efficiently.}

\noindent{\bf Dataset of user study.} We randomly select 4 apps (i.e., {Bitcoin}, {Bankdroid}, {ConnectBot}, and {Vespucci}) with different number of activities (12-15 activities) from 2 categories (i.e., finance, tool), which are hosted on Google Play Store. 
Each category contains two apps, and we ask participants to explore each app to finish the assigned tasks. 

\noindent{\bf Participant recruitment.} 
\revise{We recruit 12 people including 2 professors, 2 postdocs, and 6 Ph.D students from our university and 2 industry staff from local companies to participate in the experiment via word-of-mouth. All of the recruited participants have used Android devices for more than one year, and participated in Android related research topics. Detailed distribution is shown in Fig.~\ref{fig:distribution}.}
They never use these apps before. \revise{They come from different countries, 1 from USA, 4 from China, 4 from European countries (e.g., Spain, Germany), and 3 from Singapore.} The participants receive a \$10 shopping coupon as a compensation of their time.

\noindent{\bf Experiment procedures.} We installed the 4 apps on an Android device (Nexus 5 with Android 8.0). The experiment started with a brief introduction to the tasks. We explained and went through all the features we want them to use within the apps and asked each participant to explore and review the 4 apps separately to finish the tasks below. Note that for each category, each participant explored one app with \tool, and the other without \tool. To avoid potential bias, the order of app category, and the order of using \tool or not using are rotated based on the Latin Square~\cite{winer1962statistical}. This setup ensures that each app is explored by multiple participants with different development experience. We told each participant to complete the task with the given apps: manually explore as many functionalities of the apps as possible in 10 minutes, which is far longer than the typical average app session (71.56 seconds)~\cite{bohmer2011falling}, and understand the app functionalities with \tool; After the exploration, participants were asked to rate their satisfactoriness in exploration (on the 5-point likert scale with 1 being least satisfied and 5 being most satisfied). All participants carried out experiments independently without any discussions with each other. After performing the task, they were required to write some comments about our tool.

\begin{table}\footnotesize
    \caption{User study results of app exploration and review. $*$ denotes $p$ $<$ 0.01 and $**$ denotes $p$ $<$ 0.05.}
    \centering
    \begin{tabular}{l c c}
			\hline
			{\bf Metrics} & \begin{tabular}[c]{@{}c@{}}{\bf Manual} {\bf Exploration}\end{tabular} & 
			\begin{tabular}[c]{@{}c@{}}{\bf StoryDistiller}\end{tabular} \\ \hline
			\begin{tabular}[c]{@{}c@{}}{\bf Avg. Time (min)}\end{tabular} & 5.47 & 2.85$^*$ \\
			\begin{tabular}[c]{@{}c@{}}{\bf Avg. Coverage}\end{tabular} & 39.06\% & 88.30\%$^*$ \\
			{\bf Satisfactoriness (1-5)} & 3.99 & 4.48$^*$$^*$ \\ \hline
	\end{tabular}
	\label{tbl:study1}
\end{table}

\vspace{1mm}
\noindent{\bf Experiment results.} As displayed in Table~\ref{tbl:study1}, the average activity coverage of manual exploration is quite low (i.e., 39.06\%), showing the difficulty in exploring app functionalities thoroughly by manual exploration. However, the participants' satisfactoriness of completeness of exploration is high (i.e., 3.99 on average). It indicates that the development teams sometimes are not aware that they miss many features when exploring others' apps.
Such blind confidence and neglection may further negatively influence their strategy or decision in developing their own apps. 
\revise{Compared with manual exploration, \tool achieves over 2 times more activity coverage (\textbf{88.30\%} vs. 86.50\% in \oldtool) with less time cost (\textbf{2.85 minutes} on average vs. \revise{2.5 minutes in \oldtool}) to help understand the app functionalities. According to the \rTwo{participants}' feedback, the average satisfactoriness of \tool is \textbf{4.48} (vs. 4.40 in \oldtool), which represents the usefulness of helping participants explore and understand app functionalities.} To understand the significance of the differences between without and with \tool, we carry out the Mann-Whitney U test~\cite{utest}, which is designed for small samples. The result in Table~\ref{tbl:study1} is significant with $p$-value $<$ 0.01 or $p$-value $<$ 0.05.

\section{Dataset and Possible Applications}\label{sec:rq3}
\revise{As aforementioned, \tool is a fundamental tool which constructs a multi-dimension dataset (e.g., app storyboards and UI components). Such a rich dataset can be used to expand the horizon of current mobile app research. In this section, we discuss several application scenarios by leveraging this dataset.}

\subsection{UI Design Recommendation and Layout Code Generation}
Developing the GUI of a mobile application involves two steps, i.e., UI design and implementation. Designing a UI focuses on proper user interaction and visual effects, while implementing a UI focuses on making the UI work as designed with proper layouts and widgets of a GUI framework. For the tasks of UI design recommendation~\cite{bernal2019guigle} and layout code generation~\cite{chen2018ui}, our dataset provides a large set of diverse UI pages, as well as the corresponding layout code. The diversity of the collected data depends on \tool's ability of thoroughly exploring apps' UI pages. Additionally, it is crucial to provide real UI pages for the UI design recommendation task. Based on the results of \emph{ATG extraction} (\S~\ref{subsec:rq1}) and \emph{UI page rendering} (\S~\ref{subsec:rq2}), \tool is able to obtain a high activity coverage compared with dynamic testing tools and a high successful rate of UI page rendering. Moreover, the rendered UI pages are almost same as the real ones that users would observe.

\revise{The UI pages with attributes in our dataset can assist both UI designers and developers.} Such a dataset bridges the gap across the abstract activities (text), UI pages (image) and detailed layout code (i.e., activity $\rightarrow$ UI page $\rightarrow$ layout code) so that they can be searched as a whole. Due to such mapping relation, UI/UX designers can directly use keywords (e.g., ``Login'' and ``Search'') to search for the UI images by matching the activity name of the UI in our dataset. The searched images can be used for inspiring their own UI design. The UI developers can also benefit from searching our dataset for UI implementation. For another application scenario, given the UI design image from designers, developers can search for the similar UIs in our dataset by computing the image similarity. As each UI page in our dataset is also associated with corresponding run-time UI code, developers can choose the most related UI page in the candidate list and then customize the UI code for their own purpose to implement the given UI design.

Additionally, by training a neural machine translator, we are able to translate a UI design image to a GUI skeleton. Chen et al.~\cite{chen2018ui} collected the training data based on the dynamic testing tool, {Stoat}. However, according to the experimental results of ATG generation, we find that \tool covers 2 times more activities than {Stoat} with less time. Consequently, the results are limited to the diversity of the training data used in~\cite{chen2018ui}. Our constructed dataset of UI pages are more comprehensive with diverse UI designs.

\subsection{UI Component Recommendation}
UI component sharing provides an opportunity to learn about GUI designs, gain design inspiration and understand design trend~\cite{chen2019gallery}. To enable the recommendation task of GUI components, \revise{our dataset collected a large number of separate UI components (e.g., ``Button'') together with their attributes and the corresponding bank-end design code.} Based on them, we highlight some typical tasks or potential application scenarios as follows. (1) Alice aims to design a social media app and wants to decide the style of the buttons so that it can fit for the theme of such a social media app. With the constructed dataset, she can search for hundreds of buttons to get inspirations. According to the candidates returned by the dataset, she can choose the most attractive one as the final decision for her own apps. (2) Apart from the style of UI component, the size and color are also provided to Alice. Therefore, she may observe that social apps usually use larger size with bright color buttons for most social media apps. (3) Based on the results of multiple-time searching, Alice may also understand the design trend of UI components in one app category, which is also helpful for developing apps in specific categories.

\subsection{Code Search} 
When developers implement their own apps, aiming to ensure the competitive edge in the markets, they usually attempt to get inspirations from the similar components (e.g., Activity) implemented in other apps, because the components with the same semantic name have a great probability to own similar logic and architectures (e.g., method hierarchy). \revise{To enable such a code search task, our constructed dataset also collects the logic code with Activity names.} Firstly, we divide the apps based on their app categories, such as finance, social media, and news since the apps in the same category would contain more common features. Secondly, we store the activities if they have the same semantic activity name, such as LoginActivity, RegActivity, AboutActivity, and EditActivity.

For example, Bob is a junior app developer. For the login activity, he may only implement the basic logic, i.e., collect user' inputs and validate whether the inputs are consistent with the information stored in the server or the database. With the help of our constructed dataset, he can search for the similar implementation by the same Activity name, i.e., \textit{LoginActivity} or just \textit{Login}. After searching, he would note that he should also validate the format before collecting the users' inputs, which is a typical specification. In this case, the logic code with same name could help to improve the quality of their own apps and customize more interesting features.

\subsection{\tool for App Testing}
\noindent \textbf{App GUI testing.} \revise{According to many previous studies~\cite{choudhary2015automated,zeng2016automated}, there are only about 40\% activity coverage for most dynamic GUI app testing tools such as Monkey and Stoat, mainly due to lack of improper user input complex constraints. 
Thanks to the relatively complete ATG constructed by \tool, we can leverage it to explore more activities and enhance the exploration capability of transition-based dynamic testing tools. For example, when apps are under testing by using Monkey, we can differentiate the transitions that are never explored by Monkey by comparing the transitions and covered activities. For the uncovered transitions, based on our ATG, we can directly launch the target activities and make the testing tool start to explore from this new state (using their own exploration strategy) to explore more state and detect more bugs.}

\vspace{1mm}
\noindent \textbf{App regression testing.} Reusing test cases is useful to improve the efficiency of regression testing for Android apps~\cite{rothermel2001prioritizing}. 
\revise{\tool can help guide app regression testing by identifying the ATG and UI components that have been modified. Note that, different versions of a single app have many common functionalities, which means most of the UI pages in the newer version are the same as the previous version. The ATGs of different versions can be easily used to demonstrate the common functionalities. Meanwhile, \tool stores the mapping relation between UI page and the corresponding layout code, therefore, analyzers can obtain the modified UI components by analyzing the differences of layout code, and further update the test cases accordingly.}
In this scenario, most of the test cases can be reused, and the modified components can be identified effectively to guide test case update for regression testing.

\section{Limitations}\label{sec:limitation}
In this section, we discuss the limitations of \oldtool.

\noindent \textbf{Incomplete features due to the underlying tools.} The inputs of UI page rendering are extracted from static analysis based on {Soot}, but some files failed to be transformed, and the call graphs can still be incomplete. As for the closed-source apps, {jadx} is used to decompile apk to Java code. However, some Java files failed to be decompiled, which affects the analysis results of UI page rendering. But according to our observation, these cases rarely appear in the real apps. Besides, as the activities spawned by other components (e.g., Broadcast Receiver) can only be dynamically loaded, our static-analysis based approach cannot deal with them.

\vspace{1mm}
\noindent \textbf{Failures in UI page rendering.}
Although \tool achieves \textasciitilde80\% launch ratio of activities for each app on average, some UI pages still cannot be rendered successfully due to several errors. (1) Some activities require valid authentication information to launch, that is, they will check whether the current state owns valid authentication (e.g., successful login in) before rendering the page, if the activity tries to be launched without valid authentication, it may redirect to the sign-in or sign-up page. Such scenario is an open challenge in Android app testing, unless the testers provide the login information before hand to enable the login process, then the app can continue explore the pages that require valid authentication. Thus, \tool would fail to render such kind of activities. (2) Although we provide the required ICC data as the activity launching parameters, some activities still need to load other required data from local storage (e.g., SharedPreference, SQLite Database) or remote servers. \tool cannot provide this kind of required data so far, causing 	failures when launching this kind of activities.

\vspace{1mm}
\noindent \textbf{\rTwo{Incomplete activity presentations due to fragments.}}
\rTwo{As aforementioned in the paper, an activity may have multiple fragments in practice. 
First of all, it is possible to define in the static layout file of an activity that it contains fragments (i.e., static binding), and fragments are treated as views to render the activity. While developers can also choose to bind the fragments (e.g., add, delete, and replace) of an activity at runtime (i.e., dynamic binding). 
As for the current version of StoryDistiller, it only records one UI page per activity with static fragments. If the current activity uses the static binding method to bind fragments, StoryDistiller can leverage the proposed hybrid method to render the activity with fragments.
However, if the fragments are integrated into the activity at runtime triggered by users or specific operations, StoryDistiller cannot record the changes for different fragments in one activity so far.}

\section{Related Work}\label{sec:relatedwork}
\noindent{\bf Assist Android development}. GUI provides a visual bridge between apps and users through which they can interact with each other. Developing the GUI of a mobile app involves two separate but related activities: design the UI and implement the UI. To assist UI implementation, Nguyen and Csallner~\cite{nguyen2015reverse} reverse-engineer the UI screenshots by image processing techniques. More powerful deep-learning based algorithms~\cite{chen2018ui, beltramelli2018pix2code, moran2018machine} are further proposed to leverage the existing big data of Android apps. Retrieval-based methods~\cite{reiss2018seeking, behrang2018guifetch} are also used to develop the user interfaces. Reiss~\cite{reiss2018seeking} parses the sketch into structured queries to search related UIs of Java-based desktop software in the database.

Different from the UI implementation studies, our study focuses more on the generation of app storyboard which not only contains the UI code, but also the transitions among the UIs. In addition, the UI code generated in prior work~\cite{nguyen2015reverse, chen2018ui, beltramelli2018pix2code, moran2018machine} is all static layout, which conflicts with our observation in Section~\ref{sec:background} that developers often write Java code to dynamically render the UI. In our work, we provide developers with the original UI code (no matter static code, dynamic code, or hybrid) for each screen. Such real code makes developers more easy to customize the UIs for their own needs. Apart from the UI implementation, some studies also recommend UI design~\cite{chen2019gallery} and explore issues between UI design and its implementation. Moran et al~\cite{moran2018automated} check whether the UI implementation violates the original UI design by comparing the image similarity with computer vision techniques. They further detect and summarize GUI changes in evolving mobile apps. They rely on the dynamically running apps for collecting UI screenshots, and that is time-consuming and leads to low coverage of the app. In contrast, our method can extract most UI pages of the app statically, so it can complement with these studies for related tasks.

{GUIfectch}~\cite{behrang2018guifetch} customizes Reiss's method~\cite{reiss2018seeking} into Android app UI search by considering the transitions between UIs. It can also extract UI screenshots with corresponding transitions, but our work is different from theirs in two aspects. First, their model can only deal with open-source apps, while ours can also reverse-engineer the closed-source apps, hence leading to more generality and flexibility. On the other hand, {GUIfectch} is much more heavy-weight than our static-analysis based approach, as it relies on both static analysis for UI code extraction and dynamic analysis for transition extraction. In addition, dynamically running the app usually cannot cover all screens \revised{like Stoat}, leading to the loss of information.

\noindent{\bf Assist app comprehension by reverse engineering.} The process of reverse engineering of Android apps is that researchers rely on the state-of-the-art tools (e.g., {Apktool}~\cite{apktool}, {Androguard}~\cite{androguard}, {dex2jar}~\cite{dex2jar}, {Soot}~\cite{soot}) for decompiling an {APK} to intermediate language (e.g., {smali}, {jimple}) or Java code. Android reverse engineering is usually used to understand and analyze apps~\cite{understand}. It also can be used to extract features for Android malware detection~\cite{chen2016stormdroid}. However, reverse engineering only has the basic functionality for code review. Different from the general reverse engineering with plain decompiled code, our work extract more abstract representations, i.e., storyboard of each app to give the overview of app functionalities and mappings between the UI page and the corresponding layout code. Such storyboard can directly help product manager and designers who are of no technical expertise to understand competitor apps.

\noindent{\bf Assist Android app analysis.} Many static analysis techniques~\cite{azim2013targeted,octeau2013effective,octeau2015composite,arzt2014flowdroid,li2015iccta,fan18,fan18efficiently,chen2018mobile,chen2018ausera} have been proposed for Android apps. A$^3$E provides two strategies, targeted and depth-first exploration, for systematic testing of Android apps~\cite{azim2013targeted}. It also extracts static activity transition graphs for automatically generated test cases. Apart from the target of Android testing, we extract activity transition graphs to identify and systematically explore the storyboard of Android apps. Epicc is the first work to extract component communication~\cite{octeau2013effective}, and it determines most Intent attributes to component matching. {ICC}~\cite{octeau2015composite} significantly outperforms Epicc on the extraction ability of inter-component communication by utilizing the solver for MVC problems based on the proposed {COAL} language. FlowDroid~\cite{arzt2014flowdroid} and IccTA~\cite{li2015iccta} extract call graphs based on {Soot} for data-flow analysis for detecting data leakage and malicious behaviors~\cite{chen2016stormdroid, fan2016poster,chen2016towards,chen2018automated,chen2019poison,fakeapp, chen2018ausera, chen2018mobile, chen2019ausera}. Liu et al.~\cite{liu2016understanding} utilized program analysis to understand the patterns that cause functional and nonfunctional issues and proposed a static analysis tool to detect two most common patterns of wake lock misuses. Wei et al.~\cite{wei2017oasis} combined program analysis and NLP techniques to prioritize Lint warnings by leveraging app user reviews. Dong et al.~\cite{dongicse2020} proposed time-travel testing for Android apps that can transit to the state it explored before when needed. Wei et al.~\cite{wei2019pivot} proposed an approach that automatically learns API-device correlations of compatibility issues induced by fragmentation from existing Android apps. 
\rTwo{Yan et al.~\cite{yanicse2020} proposed multi-entry testing for Android apps by analyzing the constraints for launching an activity and the solved constraints are used to launch the activity through a third-party app.
They did not focus on ATG construction, instead, they focused on the construction of Activity Launching Models (ALM) by a static method (i.e., starting with a coarse-grained ATG mentioned in their paper). By contrast, extracting a relatively comprehensive ATG is one of the most important goals in our work, {and we not only statically extract the transitions between activities, fragments, and inner classes, but also dynamically augment ATG to construct a comprehensive graph}. In terms of dynamic exploration, their goal is to adjust the weights of their Activity Launching Context (ALC) dynamically to explore apps and find bugs {by leveraging their constructed Activity Launching Model}. Instead, our goal is to augment the transition graph extracted by the pure static method in the previous work~\cite{chen2019storydroid} {by traversing the actionable components in the UI page to explore as many transitions as possible}. As for the activity launching, their method required to build a dummy app to launch activities due to the limitation of launching via adb, while our work addresses this problem by instrumenting the app, thereby can launch the activity directly from the console via adb instead of a dummy app used in~\cite{yanicse2020}.}
Compared with them, we provide another novel solution to assist Android app testing, i.e., reveal the relations between different components together with rich attributes to help understand the semantic and functionality of apps.

\section{Conclusion}\label{sec:conclusion}
In this paper, we propose \tool, a system to distill visualized storyboards of Android apps with rich features by extracting relatively complete ATG and rendering UI pages dynamically with the help of the extracted ICC data. Such a storyboard benefits different roles (i.e., PMs, UI designers, developers, and testers) in the app development process and analysis. The extensive experiments and user study demonstrate the effectiveness and usefulness of \tool. Based on the outputs of \tool, we constructed different kinds of large-scale datasets to bridge the gap across app activities (descriptive text), UI pages (image), and implementation code (source code). In the future, we will further explore these potential applications, and also extend our approach to other platforms such as iOS apps and desktop software for more general usage.

\section*{Acknowledgments}

{We appreciate all the reviewers for their valuable comments. This work was partially supported by the National Natural Science Foundation of China (No. 62102284, 62102197).}



\bibliographystyle{IEEEtran}
\bibliography{tse}

\begin{IEEEbiography}[{\includegraphics[width=1in,height=1.25in,clip,keepaspectratio]{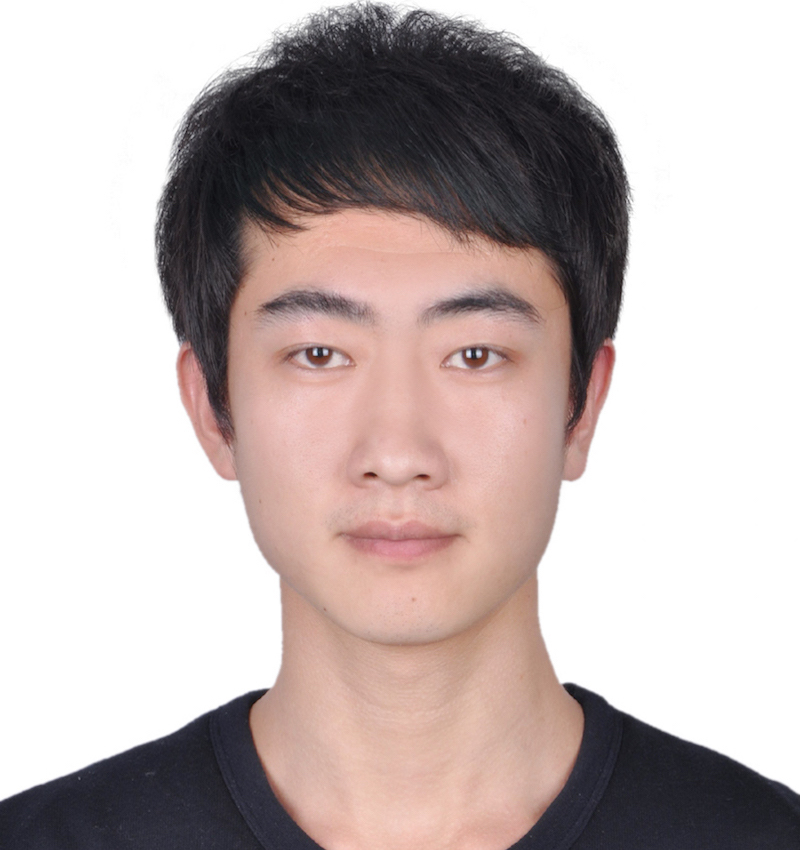}}]
{Sen Chen} (Member, IEEE) is an Associate Professor in the College of Intelligence and Computing, Tianjin University, China. Before that, he was a Research Assistant Professor in the School of Computer Science and Engineering, Nanyang Technological University, Singapore.
Previously, he was a Research Assistant of NTU from 2016 to 2019 and a Research Fellow from 2019-2020. 
He received his Ph.D. degree in Computer Science from School of Computer Science and Software Engineering, East China Normal University, China, in June 2019. 
His research focuses on Security and Software Engineering.
He has published broadly in top-tier security (IEEE S\&P, USENIX Security, CCS, IEEE TIFS, and IEEE TDSC) and software engineering venues including ICSE, FSE, ASE, ACM TOSEM, and IEEE TSE. More information is available on {\url{https://sen-chen.github.io/}.}
\end{IEEEbiography}

\begin{IEEEbiography}[{\includegraphics[width=1in,height=1.25in,clip,keepaspectratio]{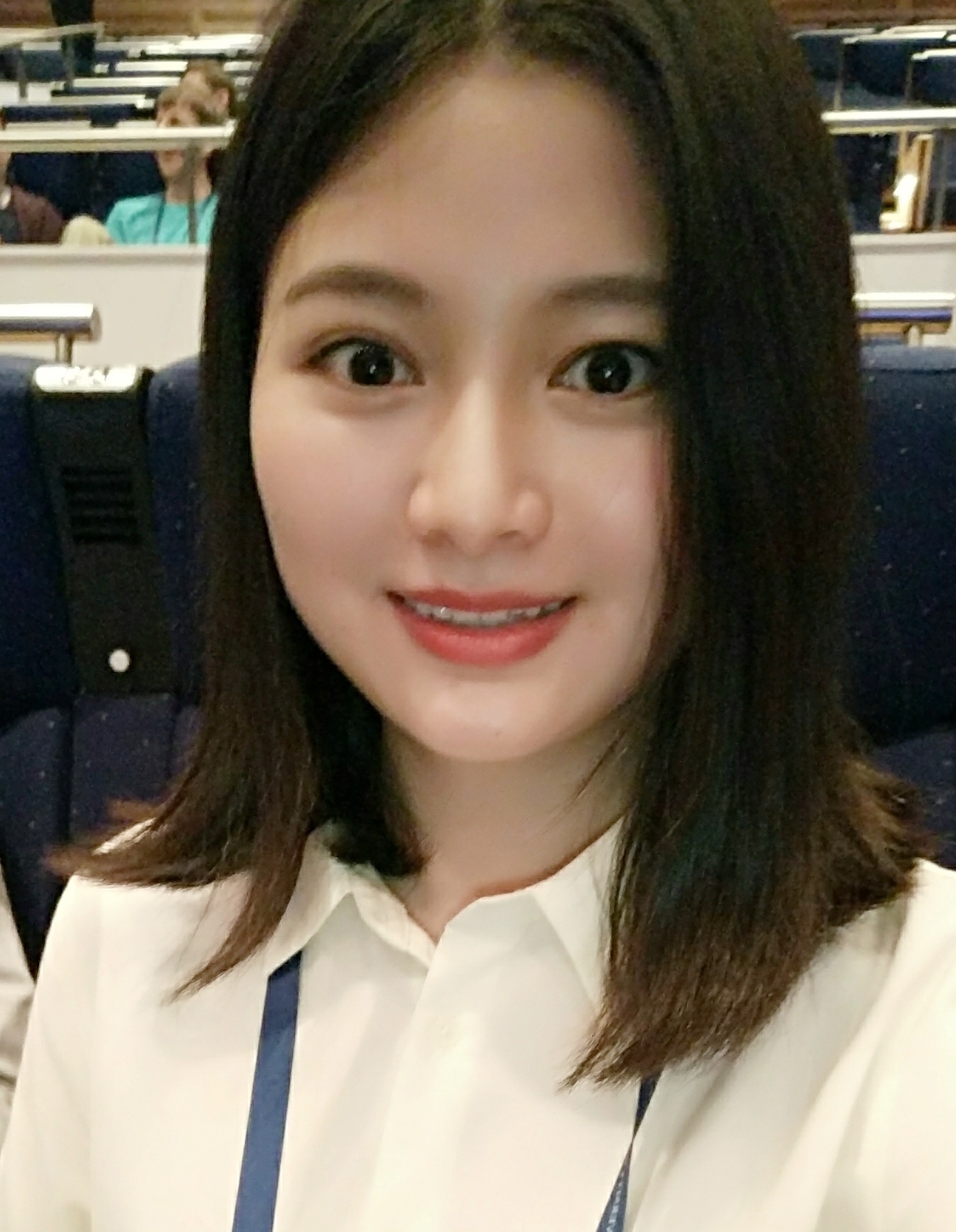}}]
{Lingling Fan} is an Associate Professor in College of Cyber Science, Nankai University, China. She received her Ph.D and BEng degrees in computer science from East China Normal University, Shanghai, China in June 2019 and June 2014, respectively. In 2017, she joined Nanyang Technological University (NTU), Singapore as a Research Assistant and then had been as a Research Fellow of NTU since 2019. Her research focuses on program analysis and testing, software security. She got two ACM SIGSOFT Distinguished Paper Awards at ICSE 2018. 
\end{IEEEbiography}

\begin{IEEEbiography}[{\includegraphics[width=1in,height=1.25in,clip,keepaspectratio]{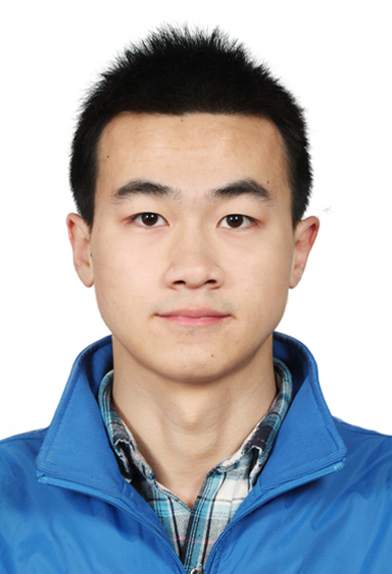}}]{Chen Chunyang}
obtained his Ph.D. degree from School of Computer Science and Engineering, Nanyang Technological University (NTU), Singapore, and bachelor's degree from Beijing University of Posts and Telecommunications (BUPT), China, June 2014. He is a lecturer (a.k.a. Assistant Professor) in Faculty of Information Technology, Monash University, Australia. His research focuses on Mining Software Repositories, Text Mining, Deep Learning, and Human Computer Interaction. 
\end{IEEEbiography}

\begin{IEEEbiography}[{\includegraphics[width=1in,height=1.25in,clip,keepaspectratio]{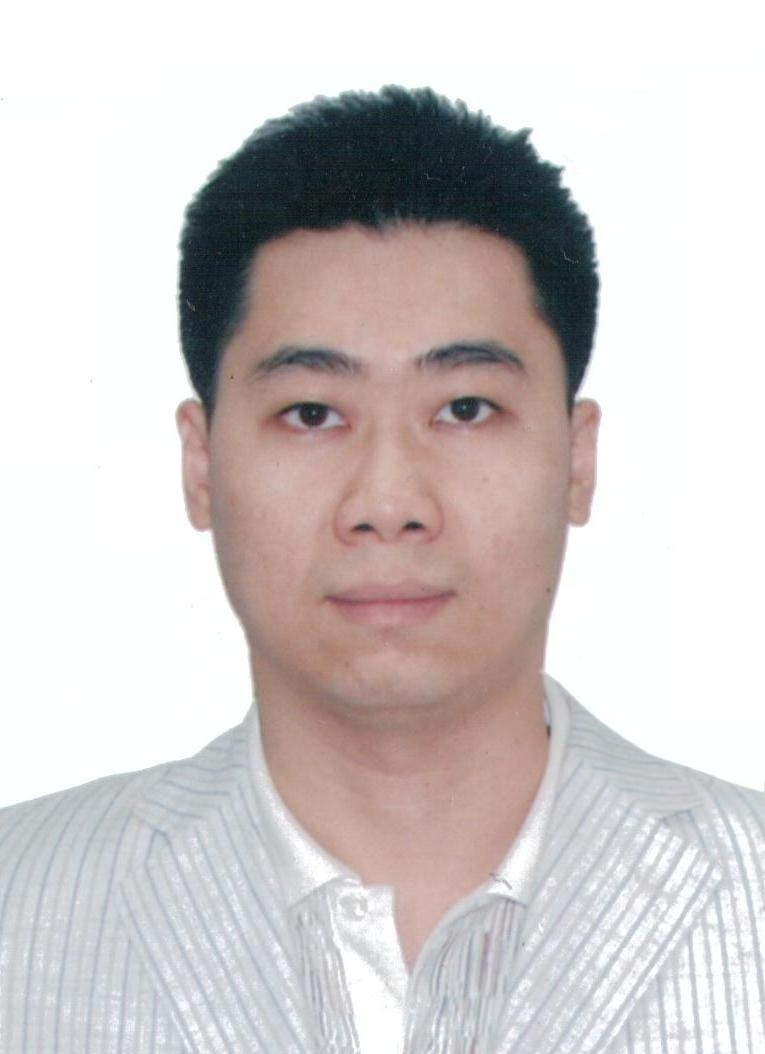}}]{Liu Yang}
graduated in 2005 with a Bachelor of Computing (Honours) in the National University of Singapore (NUS). In 2010, he obtained his PhD and started his post doctoral work in NUS, MIT and SUTD. In 2011, Dr. Liu is awarded the Temasek Research Fellowship at NUS to be the Principal Investigator in the area of Cyber Security. In 2012 fall, he joined Nanyang Technological University (NTU) as a Nanyang Assistant Professor. He is currently a full professor and the director of the cybersecurity lab in NTU.
              
He specializes in software verification, security and software engineering. His research has bridged the gap between the theory and practical usage of formal methods and program analysis to evaluate the design and implementation of software for high assurance and security. His work led to the development of a state-of-the-art model checker, Process Analysis Toolkit (PAT). By now, he has more than 300 publications and 6 best paper awards in top tier conferences and journals. With more than 20 million Singapore dollar funding support, he is leading a large research team working on the state-of-the-art software engineering and cybersecurity problems.
\end{IEEEbiography}

\end{document}